\documentclass[11pt]{article}

\usepackage[utf8]{inputenc}
\usepackage[margin=1in]{geometry}
\usepackage{amsmath,amssymb,amsthm}

\usepackage{tikz}
\usetikzlibrary{calc}
\usepackage{subcaption}
\usepackage{float}
\usepackage{graphicx}
\usepackage{hyperref}

\newtheorem{theorem}{Theorem}
\newtheorem{definition}{Definition}

\title{Maximally-Matched Mixture Decoding for Correlated Atom Loss in Quantum LDPC Codes}
\author{Jacob W. Murri$^{*}$ \\
  University of California, Los Angeles \\
  \texttt{jwmurri@math.ucla.edu}}
\date{August 21, 2026}

\begin{document}

\maketitle

\footnotetext[1]{Corresponding author}

\begin{abstract}
Heralded atom loss is a dominant noise channel in neutral atom quantum computing platforms, and losses at a two-qubit gate are strongly correlated: when one atom is lost, its gate partner is often lost too. This work shows how to exploit this correlation to construct a better decoder, focusing on $[[144,12,12]]$ Gross code syndrome extraction circuits. Given a loss readout, the possible pairings of the lost qubits according to shared gates form a graph whose maximum matchings carry almost all of the conditional probability mass. Our \emph{maximal-matching-mixture} (MMM) decoder builds a per-readout decoding hypergraph as a weighted mixture over exactly these loss scenarios. This decoding hypergraph can be used by an ordinary belief-propagation decoder. On the $[[144,12,12]]$ Gross code, MMM lowers the logical error rate by as much as ${\sim}25\times$ relative to a loss-unaware decoder and ${\sim}2\times$ relative to a loss-aware uniform-mixture baseline from Baranes et al.\ (2026).
\end{abstract}

\section{Introduction}\label{sec:intro}

Due to high error rates in physical implementations of quantum computers, large-scale quantum computation is only possible through the fault-tolerant quantum computing (FTQC) paradigm. Suppression of errors is enabled by quantum error correction (QEC) schemes, which involve encoding logical qubits in a larger number of physical qubits, along with various checks to be able to diagnose and correct errors that occur at the physical level. QEC performance depends heavily on algorithm structure, error model, and code choice for a particular quantum circuit. While surface codes have dominated experimental implementations due to their geometric locality and high thresholds, quantum low-density parity-check (qLDPC) codes promise better encoding rates and lower overhead. These codes are most naturally implemented on hardware which allows long-range interactions between qubits, such as reconfigurable neutral-atom arrays (RNAA) \cite{yang2026}.\\

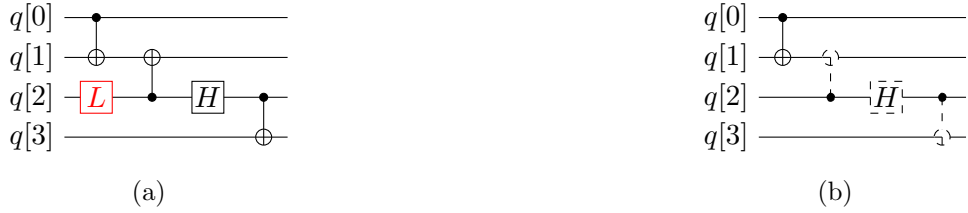
\begin{figure}[H]
\centering
\begin{subfigure}[b]{0.45\textwidth}
\centering
\begin{tikzpicture}[scale=1.000000,x=1pt,y=1pt]
\filldraw[color=white] (0.000000, -7.500000) rectangle (84.000000, 52.500000);
\draw[color=black] (0.000000,45.000000) -- (84.000000,45.000000);
\draw[color=black] (0.000000,45.000000) node[left] {$q[0]$};
\draw[color=black] (0.000000,30.000000) -- (84.000000,30.000000);
\draw[color=black] (0.000000,30.000000) node[left] {$q[1]$};
\draw[color=black] (0.000000,15.000000) -- (84.000000,15.000000);
\draw[color=black] (0.000000,15.000000) node[left] {$q[2]$};
\draw[color=black] (0.000000,0.000000) -- (84.000000,0.000000);
\draw[color=black] (0.000000,0.000000) node[left] {$q[3]$};
\begin{scope}[color=red]
\begin{scope}[color=red]
\begin{scope}
\draw[fill=white] (12.000000, 15.000000) +(-45.000000:8.485281pt and 8.485281pt) -- +(45.000000:8.485281pt and 8.485281pt) -- +(135.000000:8.485281pt and 8.485281pt) -- +(225.000000:8.485281pt and 8.485281pt) -- cycle;
\clip (12.000000, 15.000000) +(-45.000000:8.485281pt and 8.485281pt) -- +(45.000000:8.485281pt and 8.485281pt) -- +(135.000000:8.485281pt and 8.485281pt) -- +(225.000000:8.485281pt and 8.485281pt) -- cycle;
\draw (12.000000, 15.000000) node {$L$};
\end{scope}
\end{scope}
\end{scope}
\draw (12.000000,45.000000) -- (12.000000,30.000000);
\begin{scope}
\draw[fill=white] (12.000000, 30.000000) circle(3.000000pt);
\clip (12.000000, 30.000000) circle(3.000000pt);
\draw (9.000000, 30.000000) -- (15.000000, 30.000000);
\draw (12.000000, 27.000000) -- (12.000000, 33.000000);
\end{scope}
\filldraw (12.000000, 45.000000) circle(1.500000pt);
\draw (33.000000,30.000000) -- (33.000000,15.000000);
\begin{scope}
\draw[fill=white] (33.000000, 30.000000) circle(3.000000pt);
\clip (33.000000, 30.000000) circle(3.000000pt);
\draw (30.000000, 30.000000) -- (36.000000, 30.000000);
\draw (33.000000, 27.000000) -- (33.000000, 33.000000);
\end{scope}
\filldraw (33.000000, 15.000000) circle(1.500000pt);
\begin{scope}
\draw[fill=white] (54.000000, 15.000000) +(-45.000000:8.485281pt and 8.485281pt) -- +(45.000000:8.485281pt and 8.485281pt) -- +(135.000000:8.485281pt and 8.485281pt) -- +(225.000000:8.485281pt and 8.485281pt) -- cycle;
\clip (54.000000, 15.000000) +(-45.000000:8.485281pt and 8.485281pt) -- +(45.000000:8.485281pt and 8.485281pt) -- +(135.000000:8.485281pt and 8.485281pt) -- +(225.000000:8.485281pt and 8.485281pt) -- cycle;
\draw (54.000000, 15.000000) node {$H$};
\end{scope}
\draw (75.000000,15.000000) -- (75.000000,0.000000);
\begin{scope}
\draw[fill=white] (75.000000, 0.000000) circle(3.000000pt);
\clip (75.000000, 0.000000) circle(3.000000pt);
\draw (72.000000, 0.000000) -- (78.000000, 0.000000);
\draw (75.000000, -3.000000) -- (75.000000, 3.000000);
\end{scope}
\filldraw (75.000000, 15.000000) circle(1.500000pt);
\end{tikzpicture}
\caption{}
\label{fig:1a}
\end{subfigure}
\hfill
\begin{subfigure}[b]{0.45\textwidth}
\centering
\begin{tikzpicture}[scale=1.000000,x=1pt,y=1pt]
\filldraw[color=white] (0.000000, -7.500000) rectangle (78.000000, 52.500000);
\draw[color=black] (0.000000,45.000000) -- (78.000000,45.000000);
\draw[color=black] (0.000000,45.000000) node[left] {$q[0]$};
\draw[color=black] (0.000000,30.000000) -- (78.000000,30.000000);
\draw[color=black] (0.000000,30.000000) node[left] {$q[1]$};
\draw[color=black] (0.000000,15.000000) -- (78.000000,15.000000);
\draw[color=black] (0.000000,15.000000) node[left] {$q[2]$};
\draw[color=black] (0.000000,0.000000) -- (78.000000,0.000000);
\draw[color=black] (0.000000,0.000000) node[left] {$q[3]$};
\draw (9.000000,45.000000) -- (9.000000,30.000000);
\begin{scope}
\draw[fill=white] (9.000000, 30.000000) circle(3.000000pt);
\clip (9.000000, 30.000000) circle(3.000000pt);
\draw (6.000000, 30.000000) -- (12.000000, 30.000000);
\draw (9.000000, 27.000000) -- (9.000000, 33.000000);
\end{scope}
\filldraw (9.000000, 45.000000) circle(1.500000pt);
\begin{scope}[dashed]
\draw[dashed] (27.000000,30.000000) -- (27.000000,15.000000);
\begin{scope}
\draw[fill=white] (27.000000, 30.000000) circle(3.000000pt);
\clip (27.000000, 30.000000) circle(3.000000pt);
\draw (24.000000, 30.000000) -- (30.000000, 30.000000);
\draw (27.000000, 27.000000) -- (27.000000, 33.000000);
\end{scope}
\filldraw (27.000000, 15.000000) circle(1.500000pt);
\end{scope}
\begin{scope}[dashed]
\begin{scope}[dashed]
\begin{scope}
\draw[fill=white] (48.000000, 15.000000) +(-45.000000:8.485281pt and 8.485281pt) -- +(45.000000:8.485281pt and 8.485281pt) -- +(135.000000:8.485281pt and 8.485281pt) -- +(225.000000:8.485281pt and 8.485281pt) -- cycle;
\clip (48.000000, 15.000000) +(-45.000000:8.485281pt and 8.485281pt) -- +(45.000000:8.485281pt and 8.485281pt) -- +(135.000000:8.485281pt and 8.485281pt) -- +(225.000000:8.485281pt and 8.485281pt) -- cycle;
\draw (48.000000, 15.000000) node {$H$};
\end{scope}
\end{scope}
\end{scope}
\begin{scope}[dashed]
\draw[dashed] (69.000000,15.000000) -- (69.000000,0.000000);
\begin{scope}
\draw[fill=white] (69.000000, 0.000000) circle(3.000000pt);
\clip (69.000000, 0.000000) circle(3.000000pt);
\draw (66.000000, 0.000000) -- (72.000000, 0.000000);
\draw (69.000000, -3.000000) -- (69.000000, 3.000000);
\end{scope}
\filldraw (69.000000, 15.000000) circle(1.500000pt);
\end{scope}
\end{tikzpicture}
\caption{}
\label{fig:1b}
\end{subfigure}
\caption{Adapted from \cite{liu2026}. Gate stripping for atom loss. (a) A loss event at qubit $q[2]$ before the first CNOT gate. (b) The resulting circuit after gate stripping removes gates acting on the lost qubit.}
\label{fig:gate_stripping}
\end{figure}

Among the most promising platforms in quantum computing, neutral-atom quantum computers have recently seen large advances in gate fidelity and scalability. These platforms offer the advantage that connectivity is no longer limited to nearest neighbors. Physical qubits are neutral atoms held in optical tweezers, and acousto-optic deflection (AOD) is used to transport arrays of many atoms simultaneously, simplifying the implementation of QEC with qLDPC codes \cite{yang2026}. Two-qubit gates are implemented using the Rydberg interaction, which is subject to a phenomenon called atom loss, where a qubit is rendered physically inaccessible or leaks out of the computational subspace. The former typically occurs when an atom escapes from its optical trap. Atom loss is a significant error mechanism, accounting for a large proportion of all errors, and it presents several challenges. When atom loss occurs, any subsequent gates involving that atom fail to operate (see Figure~\ref{fig:gate_stripping}, adapted from \cite{liu2026}), leading to non-Pauli errors. Furthermore, atom loss persists until detection, usually at the time of measurement using state-selective readout (SSR), which can distinguish between $|0\rangle$, $|1\rangle$, or ``no atom present''. However, state-selective readout cannot determine when the loss occurred, and does not allow the loss event to be localized in general. Thus, atom loss can cause many time-correlated errors which accumulate up until the time the loss is detected, and may be termed a delayed heralded error mechanism (in contrast to an erasure error which is immediately heralded). Furthermore, because errors from gates failing to operate are non-Pauli, they do not combine and propagate linearly.\\

Recent work has developed sophisticated frameworks for incorporating loss detection into decoding algorithms. Baranes et al.~\cite{baranes2026} introduced a delayed-erasure decoder that separates background Pauli noise from loss-induced error patterns. Their approach constructs a lifecycle mixture detector error model (DEM) for each qubit by averaging over all possible loss times within that qubit's lifecycle, then combines these additively with the background noise DEM. Liu et al.~\cite{liu2026} developed the Pauli envelope framework, which characterizes the set of detector patterns consistent with a given loss configuration by inserting depolarizing noise at critical circuit locations. They demonstrated that decoding with the Pauli envelope achieves the same distance as erasure decoding while avoiding the computational overhead of explicit gate removal. Furthermore, \cite{liu2026} pioneered the use of a loss exclusivity constraint when framing the decoding problem as an integer linear program (ILP). This loss exclusivity constraint enabled lower logical error rates and higher loss distance, at the cost of computational expense and scalability, since the decoding ILP had to be solved with a general-purpose ILP solver. \\

Baranes et al.~\cite{baranes2026} focus primarily on a noise regime where loss events are uncorrelated. However, correlated loss is a dominant error mechanism for neutral atom computers \cite{erasureconversion}, and has been demonstrated to be easier to deal with than uncorrelated loss \cite{liu2026,correlatedloss}. The work of Perrin et al.~\cite{correlatedloss} focused heavily on the correlated-loss noise regime, where clever decoding strategies can effectively narrow down the possibilities for loss locations given a loss readout. Their decoder, however, is tailored to surface-code matching and does not carry over to a qLDPC decoding hypergraph; the mixture construction we introduce below is code-agnostic. \\

These prior works focused primarily on surface codes. This paper extends loss-based decoding to qLDPC codes (focusing on the Gross [[144, 12, 12]] code) and addresses multi-qubit loss scenarios where correlation between loss times is significant. Our main contribution is a conditional probability weighting scheme that exploits correlations between losses at two-qubit gates. It utilizes the fact that if two lost qubits share a two-qubit gate, the scenario where both were lost together at that gate has substantially higher probability than scenarios where they were lost independently. This scheme is not specific to any quantum error-correcting code and can be applied in general.\\

We introduce a decoder architecture called Maximal-Matching Mixture that weights loss scenarios according to their conditional probability, while maintaining efficiency by discarding extremely unlikely scenarios. It exploits the correlation structure in multi-loss scenarios by concentrating conditional probability mass on scenarios with maximal numbers of correlated qubit pairs. Given a loss readout, this decoder constructs a DEM which can be passed to a standard decoding algorithm such as belief propagation, and does not require using an expensive ILP solver.\\

This paper makes two contributions. First, we present our loss-aware decoder (the maximal-matching mixture, or MMM) which exploits the correlation structure of two-qubit-gate loss to construct a small weighted mixture of decoding hypergraphs. Compared to other loss-aware decoding approaches, this weighted mixture is cheap to build per readout and pass to a standard belief-propagation decoder. Second, we quantify the extent to which periodic data-qubit teleportation is key to making any such loss-aware decoding approach feasible on bare-ancilla syndrome extraction circuits with many rounds: it caps the number of locations where a qubit may have been lost, thus making the heralding of the loss less delayed. \\

\section{Simulating Atom Loss }\label{sec:prelim}

\subsection{Detector Error Models for Pauli Errors and Atom Loss}

Correcting errors in a quantum circuit requires determining which faults occurred based on available measurements. This is oftentimes couched in the detector formalism, where a \emph{detector} is a linear combination of measurements that are deterministic under noiseless execution of a circuit \cite{liu2026,gidney2021stim}. Typically, detectors are the $\mod 2$ sum of two consecutive stabilizer measurements of a quantum error-correcting code. Each possible fault in the circuit produces some \emph{detector pattern}, a vector of bits that specifies which detectors have flipped their parity compared to noiseless execution. Similarly, each possible fault in the circuit produces some \emph{observable pattern}, a vector of bits that specifies which logical observables have flipped compared to noiseless execution.\\

Suppose we have syndrome extraction circuit $\mathcal{C}$ which is a Clifford circuit. Then the detectors and observables are deterministic under noiseless execution and under Pauli faults. Thus, each Pauli fault corresponds to one detector pattern and one observable pattern. To be precise, suppose $\mathcal{C}$ has $n_d$ detectors and $n_\ell$ logical observables. Let $\mathcal{P}$ be the set of Pauli faults and suppose it has $n_p$ generators, i.e. $\mathcal{P} \cong \mathbb{F}_2^{n_p}$. Each fault $f \in \mathcal{P}$ corresponds to a detector pattern in $\mathbb{F}_2^{n_d}$ and an observable pattern in $\mathbb{F}_2^{n_\ell}$. Because $\mathcal{C}$ is a Clifford circuit, detector and observable patterns from distinct Pauli faults combine linearly. Thus $\mathcal{C}$ induces a linear map between Pauli faults in $\mathbb{F}_2^{n_p}$ and detector/observable patterns in $\mathbb{F}_2^{n_d} \times \mathbb{F}_2^{n_\ell}$.\\

Suppose that we have a noise model for $\mathcal{C}$ where each Pauli fault is distributed as $\text{Bernoulli}(p)$ for some $p$, and all Pauli faults are independent.  This yields a \emph{detector error model} (DEM), which consists of the linear map between Pauli faults in $\mathbb{F}_2^{n_p}$ and detector/observable patterns in $\mathbb{F}_2^{n_d} \times \mathbb{F}_2^{n_\ell}$, along with the probabilities associated with each fault. The linear map can be interpreted as a matrix or as a hypergraph where each edge is a fault and each vertex is a detector pattern (this is sometimes referred to as the \emph{decoding hypergraph}). By assuming that we can combine faults which lead to the same detector/observable patterns, we may guarantee that all detector/observable patterns are unique. Furthermore, we need not make every element of $\mathcal{P}$ an edge, it suffices to include edges corresponding to a subset which generates $\mathcal{P}$.\\

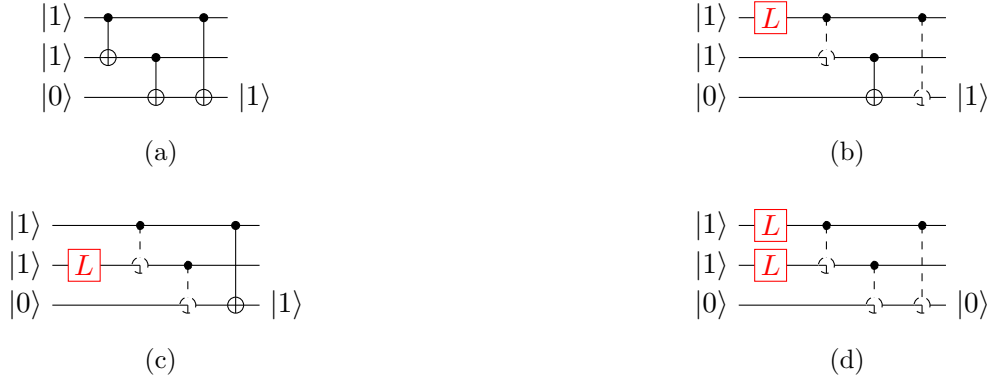
\begin{figure}[H]
\centering
\begin{subfigure}[b]{0.45\textwidth}
\centering
\begin{tikzpicture}[scale=1.000000,x=1pt,y=1pt]
\filldraw[color=white] (0.000000, -7.500000) rectangle (54.000000, 37.500000);
\draw[color=black] (0.000000,30.000000) -- (54.000000,30.000000);
\draw[color=black] (0.000000,30.000000) node[left] {$|1\rangle$};
\draw[color=black] (0.000000,15.000000) -- (54.000000,15.000000);
\draw[color=black] (0.000000,15.000000) node[left] {$|1\rangle$};
\draw[color=black] (0.000000,0.000000) -- (54.000000,0.000000);
\draw[color=black] (0.000000,0.000000) node[left] {$|0\rangle$};
\draw (9.000000,30.000000) -- (9.000000,15.000000);
\begin{scope}
\draw[fill=white] (9.000000, 15.000000) circle(3.000000pt);
\clip (9.000000, 15.000000) circle(3.000000pt);
\draw (6.000000, 15.000000) -- (12.000000, 15.000000);
\draw (9.000000, 12.000000) -- (9.000000, 18.000000);
\end{scope}
\filldraw (9.000000, 30.000000) circle(1.500000pt);
\draw (27.000000,15.000000) -- (27.000000,0.000000);
\begin{scope}
\draw[fill=white] (27.000000, 0.000000) circle(3.000000pt);
\clip (27.000000, 0.000000) circle(3.000000pt);
\draw (24.000000, 0.000000) -- (30.000000, 0.000000);
\draw (27.000000, -3.000000) -- (27.000000, 3.000000);
\end{scope}
\filldraw (27.000000, 15.000000) circle(1.500000pt);
\draw (45.000000,30.000000) -- (45.000000,0.000000);
\begin{scope}
\draw[fill=white] (45.000000, 0.000000) circle(3.000000pt);
\clip (45.000000, 0.000000) circle(3.000000pt);
\draw (42.000000, 0.000000) -- (48.000000, 0.000000);
\draw (45.000000, -3.000000) -- (45.000000, 3.000000);
\end{scope}
\filldraw (45.000000, 30.000000) circle(1.500000pt);
\draw[color=black] (54.000000,0.000000) node[right] {$|1\rangle$};
\end{tikzpicture}
\caption{}
\label{fig:2a}
\end{subfigure}
\hfill
\begin{subfigure}[b]{0.45\textwidth}
\centering
\begin{tikzpicture}[scale=1.000000,x=1pt,y=1pt]
\filldraw[color=white] (0.000000, -7.500000) rectangle (78.000000, 37.500000);
\draw[color=black] (0.000000,30.000000) -- (78.000000,30.000000);
\draw[color=black] (0.000000,30.000000) node[left] {$|1\rangle$};
\draw[color=black] (0.000000,15.000000) -- (78.000000,15.000000);
\draw[color=black] (0.000000,15.000000) node[left] {$|1\rangle$};
\draw[color=black] (0.000000,0.000000) -- (78.000000,0.000000);
\draw[color=black] (0.000000,0.000000) node[left] {$|0\rangle$};
\begin{scope}[color=red]
\begin{scope}[color=red]
\begin{scope}
\draw[fill=white] (12.000000, 30.000000) +(-45.000000:8.485281pt and 8.485281pt) -- +(45.000000:8.485281pt and 8.485281pt) -- +(135.000000:8.485281pt and 8.485281pt) -- +(225.000000:8.485281pt and 8.485281pt) -- cycle;
\clip (12.000000, 30.000000) +(-45.000000:8.485281pt and 8.485281pt) -- +(45.000000:8.485281pt and 8.485281pt) -- +(135.000000:8.485281pt and 8.485281pt) -- +(225.000000:8.485281pt and 8.485281pt) -- cycle;
\draw (12.000000, 30.000000) node {$L$};
\end{scope}
\end{scope}
\end{scope}
\begin{scope}[dashed]
\draw[dashed] (33.000000,30.000000) -- (33.000000,15.000000);
\begin{scope}
\draw[fill=white] (33.000000, 15.000000) circle(3.000000pt);
\clip (33.000000, 15.000000) circle(3.000000pt);
\draw (30.000000, 15.000000) -- (36.000000, 15.000000);
\draw (33.000000, 12.000000) -- (33.000000, 18.000000);
\end{scope}
\filldraw (33.000000, 30.000000) circle(1.500000pt);
\end{scope}
\draw (51.000000,15.000000) -- (51.000000,0.000000);
\begin{scope}
\draw[fill=white] (51.000000, 0.000000) circle(3.000000pt);
\clip (51.000000, 0.000000) circle(3.000000pt);
\draw (48.000000, 0.000000) -- (54.000000, 0.000000);
\draw (51.000000, -3.000000) -- (51.000000, 3.000000);
\end{scope}
\filldraw (51.000000, 15.000000) circle(1.500000pt);
\begin{scope}[dashed]
\draw[dashed] (69.000000,30.000000) -- (69.000000,0.000000);
\begin{scope}
\draw[fill=white] (69.000000, 0.000000) circle(3.000000pt);
\clip (69.000000, 0.000000) circle(3.000000pt);
\draw (66.000000, 0.000000) -- (72.000000, 0.000000);
\draw (69.000000, -3.000000) -- (69.000000, 3.000000);
\end{scope}
\filldraw (69.000000, 30.000000) circle(1.500000pt);
\end{scope}
\draw[color=black] (78.000000,0.000000) node[right] {$|1\rangle$};
\end{tikzpicture}
\caption{}
\label{fig:2b}
\end{subfigure}

\vspace{1em}

\begin{subfigure}[b]{0.45\textwidth}
\centering
\begin{tikzpicture}[scale=1.000000,x=1pt,y=1pt]
\filldraw[color=white] (0.000000, -7.500000) rectangle (78.000000, 37.500000);
\draw[color=black] (0.000000,30.000000) -- (78.000000,30.000000);
\draw[color=black] (0.000000,30.000000) node[left] {$|1\rangle$};
\draw[color=black] (0.000000,15.000000) -- (78.000000,15.000000);
\draw[color=black] (0.000000,15.000000) node[left] {$|1\rangle$};
\draw[color=black] (0.000000,0.000000) -- (78.000000,0.000000);
\draw[color=black] (0.000000,0.000000) node[left] {$|0\rangle$};
\begin{scope}[color=red]
\begin{scope}[color=red]
\begin{scope}
\draw[fill=white] (12.000000, 15.000000) +(-45.000000:8.485281pt and 8.485281pt) -- +(45.000000:8.485281pt and 8.485281pt) -- +(135.000000:8.485281pt and 8.485281pt) -- +(225.000000:8.485281pt and 8.485281pt) -- cycle;
\clip (12.000000, 15.000000) +(-45.000000:8.485281pt and 8.485281pt) -- +(45.000000:8.485281pt and 8.485281pt) -- +(135.000000:8.485281pt and 8.485281pt) -- +(225.000000:8.485281pt and 8.485281pt) -- cycle;
\draw (12.000000, 15.000000) node {$L$};
\end{scope}
\end{scope}
\end{scope}
\begin{scope}[dashed]
\draw[dashed] (33.000000,30.000000) -- (33.000000,15.000000);
\begin{scope}
\draw[fill=white] (33.000000, 15.000000) circle(3.000000pt);
\clip (33.000000, 15.000000) circle(3.000000pt);
\draw (30.000000, 15.000000) -- (36.000000, 15.000000);
\draw (33.000000, 12.000000) -- (33.000000, 18.000000);
\end{scope}
\filldraw (33.000000, 30.000000) circle(1.500000pt);
\end{scope}
\begin{scope}[dashed]
\draw[dashed] (51.000000,15.000000) -- (51.000000,0.000000);
\begin{scope}
\draw[fill=white] (51.000000, 0.000000) circle(3.000000pt);
\clip (51.000000, 0.000000) circle(3.000000pt);
\draw (48.000000, 0.000000) -- (54.000000, 0.000000);
\draw (51.000000, -3.000000) -- (51.000000, 3.000000);
\end{scope}
\filldraw (51.000000, 15.000000) circle(1.500000pt);
\end{scope}
\draw (69.000000,30.000000) -- (69.000000,0.000000);
\begin{scope}
\draw[fill=white] (69.000000, 0.000000) circle(3.000000pt);
\clip (69.000000, 0.000000) circle(3.000000pt);
\draw (66.000000, 0.000000) -- (72.000000, 0.000000);
\draw (69.000000, -3.000000) -- (69.000000, 3.000000);
\end{scope}
\filldraw (69.000000, 30.000000) circle(1.500000pt);
\draw[color=black] (78.000000,0.000000) node[right] {$|1\rangle$};
\end{tikzpicture}
\caption{}
\label{fig:2c}
\end{subfigure}
\hfill
\begin{subfigure}[b]{0.45\textwidth}
\centering
\begin{tikzpicture}[scale=1.000000,x=1pt,y=1pt]
\filldraw[color=white] (0.000000, -7.500000) rectangle (78.000000, 37.500000);
\draw[color=black] (0.000000,30.000000) -- (78.000000,30.000000);
\draw[color=black] (0.000000,30.000000) node[left] {$|1\rangle$};
\draw[color=black] (0.000000,15.000000) -- (78.000000,15.000000);
\draw[color=black] (0.000000,15.000000) node[left] {$|1\rangle$};
\draw[color=black] (0.000000,0.000000) -- (78.000000,0.000000);
\draw[color=black] (0.000000,0.000000) node[left] {$|0\rangle$};
\begin{scope}[color=red]
\begin{scope}[color=red]
\begin{scope}
\draw[fill=white] (12.000000, 30.000000) +(-45.000000:8.485281pt and 8.485281pt) -- +(45.000000:8.485281pt and 8.485281pt) -- +(135.000000:8.485281pt and 8.485281pt) -- +(225.000000:8.485281pt and 8.485281pt) -- cycle;
\clip (12.000000, 30.000000) +(-45.000000:8.485281pt and 8.485281pt) -- +(45.000000:8.485281pt and 8.485281pt) -- +(135.000000:8.485281pt and 8.485281pt) -- +(225.000000:8.485281pt and 8.485281pt) -- cycle;
\draw (12.000000, 30.000000) node {$L$};
\end{scope}
\end{scope}
\end{scope}
\begin{scope}[color=red]
\begin{scope}[color=red]
\begin{scope}
\draw[fill=white] (12.000000, 15.000000) +(-45.000000:8.485281pt and 8.485281pt) -- +(45.000000:8.485281pt and 8.485281pt) -- +(135.000000:8.485281pt and 8.485281pt) -- +(225.000000:8.485281pt and 8.485281pt) -- cycle;
\clip (12.000000, 15.000000) +(-45.000000:8.485281pt and 8.485281pt) -- +(45.000000:8.485281pt and 8.485281pt) -- +(135.000000:8.485281pt and 8.485281pt) -- +(225.000000:8.485281pt and 8.485281pt) -- cycle;
\draw (12.000000, 15.000000) node {$L$};
\end{scope}
\end{scope}
\end{scope}
\begin{scope}[dashed]
\draw[dashed] (33.000000,30.000000) -- (33.000000,15.000000);
\begin{scope}
\draw[fill=white] (33.000000, 15.000000) circle(3.000000pt);
\clip (33.000000, 15.000000) circle(3.000000pt);
\draw (30.000000, 15.000000) -- (36.000000, 15.000000);
\draw (33.000000, 12.000000) -- (33.000000, 18.000000);
\end{scope}
\filldraw (33.000000, 30.000000) circle(1.500000pt);
\end{scope}
\begin{scope}[dashed]
\draw[dashed] (51.000000,15.000000) -- (51.000000,0.000000);
\begin{scope}
\draw[fill=white] (51.000000, 0.000000) circle(3.000000pt);
\clip (51.000000, 0.000000) circle(3.000000pt);
\draw (48.000000, 0.000000) -- (54.000000, 0.000000);
\draw (51.000000, -3.000000) -- (51.000000, 3.000000);
\end{scope}
\filldraw (51.000000, 15.000000) circle(1.500000pt);
\end{scope}
\begin{scope}[dashed]
\draw[dashed] (69.000000,30.000000) -- (69.000000,0.000000);
\begin{scope}
\draw[fill=white] (69.000000, 0.000000) circle(3.000000pt);
\clip (69.000000, 0.000000) circle(3.000000pt);
\draw (66.000000, 0.000000) -- (72.000000, 0.000000);
\draw (69.000000, -3.000000) -- (69.000000, 3.000000);
\end{scope}
\filldraw (69.000000, 30.000000) circle(1.500000pt);
\end{scope}
\draw[color=black] (78.000000,0.000000) node[right] {$|0\rangle$};
\end{tikzpicture}
\caption{}
\label{fig:2d}
\end{subfigure}
\caption{Adapted from \cite{liu2026}: Nonlinear combination of loss events. (a) Two independent single-qubit loss events. (b) Gate stripping for loss at $q[0]$ only. (c) Gate stripping for loss at $q[1]$ only. (d) Gate stripping for both losses combined---note the nonlinear effect where gates between the two lost qubits are removed.}
\label{fig:nonlinear_loss}
\end{figure}

Atom loss events also lead to detector/observable patterns, but these patterns do not combine linearly as they do for Pauli errors (See Figure~\ref{fig:nonlinear_loss}, adapted from \cite{liu2026}). This is a significant challenge for decoding. Even if we know exactly which atoms were lost and when, there is no simple closed-form expression for the detector/observable patterns \cite{liu2026}. Constructing a DEM for an atom loss event thus requires some care.\\

\subsection{Loss Scenarios and Loss Readouts}

Let $Q$ denote the set of qubits that may be lost during circuit execution. For each qubit $q \in Q$, let $T_q = \{t_{q,1}, t_{q,2}, \ldots, t_{q,|T_q|}\}$ be the set of instruction indices where $q$ could be lost. A \emph{loss scenario} $l$ is a set of tuples $(q, t)$ where $q \in Q$ and $t \in T_q$, specifying for each lost qubit the instruction index at which it was lost (using the convention that the loss occurs immediately before gates at that index). Let $L$ denote the set of all possible loss scenarios.\\

We define the \emph{loss-resolving readout function} $\mathcal{R}: L \to 2^Q$ as in \cite{liu2026}. It maps each loss scenario to a \emph{loss readout}, or a set of qubits that were lost, discarding timing information. For a readout $r \subseteq Q$, the preimage $\mathcal{R}^{-1}(r)$ contains all loss scenarios consistent with that readout.\\

\subsection{Pauli Envelope}\label{sec:pauli-envelope}

In this section we follow the definition of Pauli envelope used in \cite{liu2026}. Begin by defining the Pauli envelope of a single-qubit loss scenario $\{(q, t)\}$ as follows (without loss of generality, we assume that our Clifford circuit $\mathcal{C}$ is composed of CZ, H, and S gates): 

\begin{definition}{Pauli Envelope of Single Atom Loss (\cite{liu2026})} Consider a single-qubit loss scenario $\{(q, t)\}$ in the noisy Clifford circuit $\mathcal{C}$ composed of CZ, H, and S gates. Let $L_{q, t} = \{I_{q,t}, X_{q,t}, Y_{q,t}, Z_{q,t}\}$ be the set of all single-qubit Pauli gates on qubit $q$ at time $t$, and define the binary sum of multiple such sets as the set of all componentwise binary sums. Let $t$ be the time of loss, let $t_1, \ldots, t_n$ be times just after each H gate acting on $q$, and let $t_m$ be the time immediately before $q$ is measured. Then the Pauli envelope of $\{(q, t)\}$ is 
\[
E = L_{q, t} \oplus L_{q, t_1} \oplus \cdots \oplus L_{q, t_n} \oplus L_{q, t_m}.
\]
\end{definition}

Note that a CX (CNOT) gate can be constructed from H gates and a CZ gate, so if our syndrome extraction circuit is constructed from CNOT gates, $t_1, \ldots, t_n$ correspond to times where $q$ switches from target to control in a CNOT gate.  \\

For a multiple-qubit loss event, the definition extends naturally: 

\begin{definition}{Pauli Envelope of Multiple Atom Loss (\cite{liu2026})} Consider a loss scenario $l = \{(q_1, t_1), \ldots (q_n, t_n) \}$ in the noisy Clifford circuit $\mathcal{C}$. The Pauli envelope of $l$ is constructed by taking the binary sum of Pauli envelopes of the constituent single-qubit events: $E(l) = \bigoplus_{i=1}^{n} E(\{(q_i, t_i)\})$. 
\end{definition}

The importance of the Pauli envelope is that the set of all detector-observable patterns resulting from a loss scenario is a subset of the set of all detector/observable patterns resulting from elements of the Pauli envelope. To make this precise, let $S(p, l) = (\mathcal{D}(p, l), \mathcal{O}(p, l)) \in \mathbb{F}_2^{n_d} \times \mathbb{F}_2^{n_\ell}$ be the detector-observable pattern resulting from a Pauli error $p$ and a loss scenario $l$, and define $S(P, L) = \{S(p, l): p \in P, l \in L\}$ for sets of Pauli errors $P$ and sets of loss scenarios $L$. Then the following holds:

\begin{theorem}{Pauli Envelope is a Superset (\cite{liu2026})}\label{thm:pauli_envelope_superset} Let $E(l)$ be the Pauli envelope of the loss scenario $l$. Then $S(p, l) \subseteq S(p \oplus E(l), \emptyset)$ for any Pauli error $p$, where $p \oplus E(l) = \{p \oplus e: e \in E(l)\}$. 
\end{theorem}

Thus, for any loss scenario, if we can correctly decode its Pauli envelope, we can guarantee success at decoding the atom loss. Note that Liu et al. only show that being able to decode \emph{all} errors in the Pauli envelope guarantees success at decoding a particular loss. This result does not depend on the distribution of detector-observable patterns in the Pauli envelope or the distribution of detector-observable patterns from loss. \\

However, for the circuits used in this work, we can say a little more. For single-qubit loss events, in most cases the Pauli envelope exactly corresponds to the detector-observable pairs generated by removing gates due to a loss event (i.e. $S(p, l) = S(p \oplus E(l), \emptyset)$) and has at most one bit of slack (i.e. $|S(p \oplus E(l), \emptyset)| \leq 2 |S(p, l)|$) for every tested case. Even in the scenarios with one bit of slack, the additional basis element contains no observable flips. Furthermore, the distribution of detector-observable pairs was found to be the same (uniform) in each case. \\

For these reasons, the logical error rate of any given decoder on a circuit with the Pauli noise $P \oplus E(l)$ is a faithful approximation of the logical error rate of the decoder when actually stripping gates according to the loss event $l$. See Appendix~\ref{sec:envelope_vs_stripping} for details. \\

Because the Pauli envelope is much easier to model in the simulation software Stim \cite{gidney2021stim} and for the above reasons, we use it to model loss scenarios in this work.\\

\section{ Syndrome Extraction Circuits}

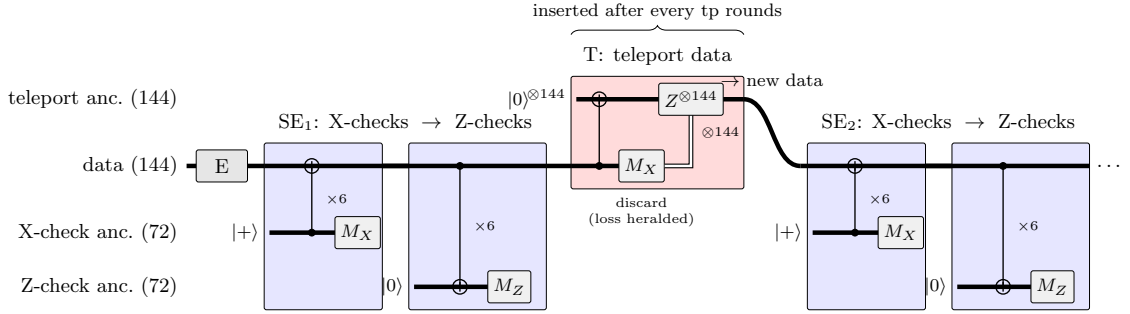
\begin{figure}[htbp]
\centering
\resizebox{\textwidth}{!}{
%
%
%
%
\usetikzlibrary{calc,decorations.pathreplacing}
\begin{tikzpicture}[x=0.55cm, y=0.65cm,
    every node/.style={font=\footnotesize},
    bundle/.style={line width=2pt},
    thinl/.style={line width=0.6pt},
    cwire/.style={double, double distance=1.1pt, line width=0.5pt},
    blk/.style={draw, fill=#1, rounded corners=1pt, inner sep=2.5pt,
                minimum height=0.5cm, align=center},
    blk/.default={black!6},
    lbl/.style={anchor=east, font=\footnotesize},
  ]
  \def\yT{5.0}    
  \def\yD{3.4}    
  \def\yAX{1.8}   
  \def\yAZ{0.5}   
  \def\seA{2.2}\def\seM{6.2}\def\seB{10.2}     
  \def\tA{10.9}\def\tB{15.8}                   
  \def\scA{17.6}\def\scM{21.6}\def\scB{25.6}   
  \def\XR{25.6}
  \node[lbl] at (0,\yT)  {teleport anc.\ $(144)$};
  \node[lbl] at (0,\yD)  {data $(144)$};
  \node[lbl] at (0,\yAX) {X-check anc.\ $(72)$};
  \node[lbl] at (0,\yAZ) {Z-check anc.\ $(72)$};
  \newcommand{\CNOT}[3]{
    \draw[thinl] (#1,#2) -- (#1,#3);
    \filldraw (#1,#2) circle(1.5pt);
    \draw[thinl] (#1,#3) circle(3.2pt);
    \draw[thinl] (#1-0.115,#3) -- (#1+0.115,#3);
    \draw[thinl] (#1,#3-0.115) -- (#1,#3+0.115);
  }
  \newcommand{\SEround}[4]{
    \draw[fill=blue!10, rounded corners=1pt]
         (#1,\yAZ-0.55) rectangle (#2-0.65,\yD+0.55);
    \draw[fill=blue!10, rounded corners=1pt]
         (#2+0.1,\yAZ-0.55) rectangle (#3,\yD+0.55);
    \draw[bundle] (#1,\yD) -- (#3,\yD);   
    \node at ({(#1+#3)/2},\yD+1.05) {#4};
    \pgfmathsetmacro\xCX{#1+1.35}\pgfmathsetmacro\xMX{#2-1.35}%
    \draw[bundle] ({#1+0.15},\yAX) -- (\xMX,\yAX);          
    \node[anchor=east,font=\scriptsize] at ({#1+0.1},\yAX) {$|{+}\rangle$};
    \CNOT{\xCX}{\yAX}{\yD}
    \node[font=\tiny,anchor=west] at ({\xCX+0.15},{(\yAX+\yD)/2}) {$\times 6$};
    \node[blk] (mx) at (\xMX,\yAX) {\scriptsize $M_X$};
    \pgfmathsetmacro\xCZ{#2+1.55}\pgfmathsetmacro\xMZ{#3-1.05}%
    \draw[bundle] ({#2+0.25},\yAZ) -- (\xMZ,\yAZ);          
    \node[anchor=east,font=\scriptsize] at ({#2+0.2},\yAZ) {$|0\rangle$};
    \CNOT{\xCZ}{\yD}{\yAZ}
    \node[font=\tiny,anchor=west] at ({\xCZ+0.15},{(\yAZ+\yD)/2}) {$\times 6$};
    \node[blk] at (\xMZ,\yAZ) {\scriptsize $M_Z$};
  }
  \draw[bundle] (0,\yD) -- (\tA,\yD);           
  \node[blk=black!10, minimum width=0.8cm] (E) at (1.0,\yD) {E};
  \SEround{\seA}{\seM}{\seB}{$\mathrm{SE}_1$: X-checks $\;\to\;$ Z-checks}
  \draw[fill=red!14, rounded corners=1pt]
       (\tA,\yD-0.55) rectangle (\tB,\yT+0.55);
  \node at ({(\tA+\tB)/2},\yT+1.05) {$\mathrm{T}$: teleport data};
  \draw[bundle] (\tA+0.15,\yT) -- (\tB,\yT);
  \node[anchor=east,font=\scriptsize] at (\tA+0.1,\yT) {$|0\rangle^{\!\otimes 144}$};
  \def\xCX{11.7}\def\xMX{12.9}\def\xZ{14.3}
  \draw[bundle] (\tA,\yD) -- (\xMX,\yD);
  \CNOT{\xCX}{\yD}{\yT}
  \node[blk] (MX) at (\xMX,\yD) {\scriptsize $M_X$};
  \node[anchor=north,font=\tiny,align=center]
       at (\xMX,\yD-0.55) {discard\\(loss heralded)};
  \draw[cwire] (MX.east) -- (\xZ,\yD) -- (\xZ,\yT-0.28);
  \node[font=\tiny,anchor=west] at (\xZ+0.05,{(\yD+\yT)/2}) {$\otimes 144$};
  \node[blk] at (\xZ,\yT) {\scriptsize $Z^{\otimes 144}$};
  \draw[bundle] (\tB,\yT) .. controls +(0.8,0) and +(-0.8,0) .. (\tB+1.6,\yD);
  \draw[bundle] (\tB+1.6,\yD) -- (\XR,\yD);
  \node[font=\scriptsize,anchor=south] at (\tB+0.8,\yT+0.1) {$\to$ new data};
  \SEround{\scA}{\scM}{\scB}{$\mathrm{SE}_2$: X-checks $\;\to\;$ Z-checks}
  \draw[decorate,decoration={brace,amplitude=4pt}]
       (\tA,\yT+1.55) -- (\tB,\yT+1.55)
       node[midway,above=3pt,font=\scriptsize]
       {inserted after every $\mathrm{tp}$ rounds};
  \node at (\XR+0.6,\yD) {$\cdots$};
\end{tikzpicture}}
\caption{A schematic of the syndrome extraction circuits used in this paper. Each wire is drawn only for the lifetime of the qubits it carries. The circuit begins with an encoder (E) to place the data qubits in the codespace. In each syndrome-extraction round, fresh $X$-check ancillas ($|{+}\rangle$) control six layers of CNOTs onto the data qubits and are measured in $X$; then fresh $Z$-check ancillas ($|0\rangle$) are the targets of six CNOT layers from the data qubits and are measured in $Z$. After every $\mathrm{tp}$ rounds, each data qubit is teleported onto a fresh ancilla (top bundle) by a CNOT, an $X$-basis measurement of the old data qubit (where loss of that qubit would be discovered) and a classically-controlled $Z$ on the ancilla, which then becomes the new data qubit. More frequent teleportation makes the circuit more loss-tolerant (for more detail see Section \ref{sec:teleportation}). }
\label{fig:circuit_overview}
\end{figure}

This work considers syndrome extraction circuits which could be used for logical quantum memory. All circuits use the Gross code, which is a [[144, 12, 12]] bivariate bicyclic CSS code. We construct a bare-ancilla syndrome extraction schedule where data-to-ancilla CNOTs are scheduled by edge-coloring the Tanner graph. Each syndrome extraction circuit consists of an encoder, which places the 144 data qubits into the codespace, followed by $n_r$ noisy rounds of $X$ and $Z$ stabilizer measurements and one final noiseless round to make sure that the decoder can propose corrections which return the data to the codespace. A schematic of these circuits is given in Figure \ref{fig:circuit_overview}.\\

\subsection{Teleportation}\label{sec:teleportation} 

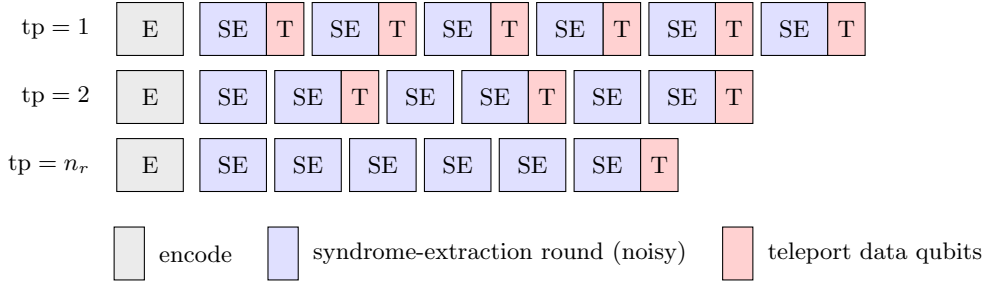
\begin{figure}[htbp]
\centering
%
\usetikzlibrary{calc}
\begin{tikzpicture}[x=0.55cm, y=0.9cm,
    every node/.style={font=\footnotesize},
    enc/.style={draw, fill=black!8, minimum height=0.7cm,
                minimum width=0.55cm*1.6, inner sep=0pt},
    se/.style ={draw, fill=blue!12, minimum height=0.7cm,
                minimum width=0.55cm*1.6, inner sep=0pt},
    tp/.style ={draw, fill=red!18, minimum height=0.7cm,
                minimum width=0.55cm*0.9, inner sep=0pt},
  ]
  \def\NR{6}
  \newcommand{\row}[3]{%
    \node[anchor=east] at (-0.4, #1) {#3};
    \node[enc] at (0.8, #1) {E};
    \pgfmathsetmacro{\x}{2.0}
    \foreach \r in {1,...,\NR} {%
      \node[se] at (\x+0.8, #1) {SE};
      \pgfmathparse{int(mod(\r, #2)==0 ? 1 : 0)}
      \ifnum\pgfmathresult=1
        \node[tp] at (\x+2.05, #1) {T};
        \pgfmathsetmacro{\x}{\x + 2.7}
      \else
        \pgfmathsetmacro{\x}{\x + 1.8}
      \fi
      \global\let\x\x
    }%
  }
  \row{2}{1}{$\mathrm{tp}=1$}
  \row{1}{2}{$\mathrm{tp}=2$}
  \row{0}{\NR}{$\mathrm{tp}=n_r$}
  \begin{scope}[shift={(0,-1.3)}]
    \node[enc,minimum width=0.4cm] (le) at (0.3,0) {};
    \node[anchor=west] (lel) at (le.east) {\,encode};
    \node[se,minimum width=0.4cm,anchor=west] (ls) at ($(lel.east)+(0.6,0)$) {};
    \node[anchor=west] (lsl) at (ls.east)
      {\,syndrome-extraction round (noisy)};
    \node[tp,minimum width=0.4cm,anchor=west] (lt) at ($(lsl.east)+(0.6,0)$) {};
    \node[anchor=west] at (lt.east) {\,teleport data qubits};
  \end{scope}
\end{tikzpicture}
\caption{A high-level timeline of the syndrome extraction circuits used in this paper, with emphasis on the placement of teleportation units which are used to swap out the physical atoms used for data qubits. In the $\mathrm{tp} = 1$ case, teleportation is done after every syndrome extraction round; in the $\mathrm{tp} = 2$ case it is only done after every other round, and in the limiting $\mathrm{tp} = n_r$ case, teleportation is only done after the last noisy round of syndrome extraction. Note that we assume that $n_r$, the number of noisy syndrome extraction rounds, is always divisible by $\mathrm{tp}$, so that data qubits are read out after the end of the last noisy syndrome extraction round. }
\label{fig:timeline}
\end{figure}

When two-qubit gates are subject to loss, logical error rates suffer when physical qubits participate in many gates before undergoing measurement and state-selective readout \cite{baranes2026}. To analyze this, the syndrome extraction circuits we consider include a teleportation of the data qubits after a certain number of syndrome extraction rounds (this is called the teleportation period or ``tp''). Each data qubit is teleported separately to a fresh ancilla qubit. Figure \ref{fig:timeline} gives a high-level overview of the circuit timeline and when teleportations are used depending on the teleportation period. \\

\begin{figure}[htbp]
\centering
\begin{tikzpicture}[scale=1.000000,x=1pt,y=1pt]
\filldraw[color=white] (0.000000, -7.500000) rectangle (84.000000, 22.500000);
\draw[color=black] (0.000000,15.000000) -- (30.000000,15.000000);
\draw[color=black] (30.000000,14.500000) -- (84.000000,14.500000);
\draw[color=black] (30.000000,15.500000) -- (84.000000,15.500000);
\draw[color=black] (0.000000,15.000000) node[left] {$d_i$};
\draw[color=black] (0.000000,0.000000) -- (84.000000,0.000000);
\draw[color=black] (0.000000,0.000000) node[left] {$a_i{=}|0\rangle$};
\draw (9.000000,15.000000) -- (9.000000,0.000000);
\begin{scope}
\draw[fill=white] (9.000000, 0.000000) circle(3.000000pt);
\clip (9.000000, 0.000000) circle(3.000000pt);
\draw (6.000000, 0.000000) -- (12.000000, 0.000000);
\draw (9.000000, -3.000000) -- (9.000000, 3.000000);
\end{scope}
\filldraw (9.000000, 15.000000) circle(1.500000pt);
\draw[fill=white] (24.000000, 11.000000) -- (32.000000,11.000000) arc (-90:90:4.000000pt) -- (24.000000,19.000000) -- cycle;
\draw (30.000000, 15.000000) node {$X$};
\draw (53.500000,15.000000) -- (53.500000,0.000000);
\draw (54.500000,15.000000) -- (54.500000,0.000000);
\begin{scope}
\draw[fill=white] (54.000000, -0.000000) +(-45.000000:8.485281pt and 8.485281pt) -- +(45.000000:8.485281pt and 8.485281pt) -- +(135.000000:8.485281pt and 8.485281pt) -- +(225.000000:8.485281pt and 8.485281pt) -- cycle;
\clip (54.000000, -0.000000) +(-45.000000:8.485281pt and 8.485281pt) -- +(45.000000:8.485281pt and 8.485281pt) -- +(135.000000:8.485281pt and 8.485281pt) -- +(225.000000:8.485281pt and 8.485281pt) -- cycle;
\draw (54.000000, -0.000000) node {$Z$};
\end{scope}
\filldraw (54.000000, 15.000000) circle(1.500000pt);
\draw (74.500000,15.000000) -- (74.500000,0.000000);
\draw (75.500000,15.000000) -- (75.500000,0.000000);
\begin{scope}
\draw (72.878680, 12.878680) -- (77.121320, 17.121320);
\draw (72.878680, 17.121320) -- (77.121320, 12.878680);
\end{scope}
\begin{scope}
\draw (72.878680, -2.121320) -- (77.121320, 2.121320);
\draw (72.878680, 2.121320) -- (77.121320, -2.121320);
\end{scope}
\end{tikzpicture}
\caption{Loss detection via teleportation. A fresh ancilla qubit is reset to $|0\rangle$, entangled with the data qubit via CNOT, the data qubit is measured in the X basis, and a classical Z correction is applied to the ancilla based on the measurement outcome. If the data qubit has been lost, the measurement will reveal its absence. Finally, the qubits are relabeled (swapped) so that subsequent operations on the logical data qubit address the physical ancilla. }
\label{fig:teleportation_gadget}
\end{figure}
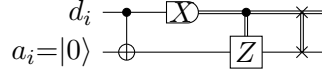

The teleportation gadget used to accomplish this is illustrated in Figure \ref{fig:teleportation_gadget}. It is sometimes called a loss-detection unit (LDU) because of its capability to detect a loss event \cite{correlatedloss}. It consists of a fresh ancilla qubit starting in the $|0\rangle$ state, which is entangled with the data qubit via CNOT. After the data qubit is measured in the X-basis, a measurement-dependent correction is applied to the ancilla. Then all subsequent operations use the new qubit. This means that the physical atom which acts as the data qubit has changed. Thus, each teleportation ends the lifetime of a qubit, and each lost data qubit is out of the circuit for at most $\mathrm{tp}$ rounds before its loss is detected. Inserting more teleportation gadgets decreases the lifetimes of physical qubits, at the cost of some noise due to the presence of an extra CNOT. \\ 

\subsection{Noise Model}

We consider quantum circuits subject to background circuit-level Pauli noise and delayed heralded atom loss. For a given physical error rate $p$ and a loss fraction $\alpha$, two-qubit gates in the circuit experience atom loss with probability $p \alpha$ and depolarizing noise with probability $p(1-\alpha)$. This latter mechanism consists of applying the Pauli gates $X$, $Y$, or $Z$ each with probability $p(1-\alpha)/3$. \\

We assume that atom loss is detected upon measurement via state-selective readout, and replenishment of lost atoms is performed by preparing new atoms initialized in the $|0\rangle$ state. Note that atom loss is distinct from an erasure error since it is not immediately heralded, and its effect is that subsequent gates operating on a lost atom do not occur. When we say that an atom loss occurs ``at'' a gate, we take this to mean that the loss occurs immediately before the attempted operation of the gate, such that the gate does not actually operate. \\

The loss probability model is based on independent mechanisms at each two-qubit gate. Let $p_\text{loss} = p \alpha$ be the probability that at least one qubit is lost during a two-qubit gate, and let $\eta$ be the conditional probability that both qubits are lost, given that a loss occurs. This yields three types of loss mechanisms for a two-qubit gate on qubits $\{a, b\}$ at time $t$: 
\begin{itemize}
\item Both qubits lost at gate $(a,b,t)$: probability $p_{\text{both}} = p_\text{loss} \eta$
\item Only qubit $a$ lost at gate $(a,b,t)$: probability $p_{\text{single}} = p_\text{loss}(1-\eta)/2$
\item Only qubit $b$ lost at gate $(a,b,t)$: probability $p_{\text{single}} = p_\text{loss}(1-\eta)/2$
\end{itemize}

Note that these probabilities are conditional upon the qubits $\{a, b\}$ not already being lost at time $t$. To be precise, using $p_{\text{both}}$ and $p_{\text{single}}$ as defined above, we have that 
\[
P( (a, t) \text{ occurs } | a \text{ not lost before } t ) = p_\text{both} + p_\text{single} 
\]
and similarly for $b$. Furthermore,
\[
P( \{(a, t), (b, t)\} \text{ occurs } | a, b \text{ not lost before } t ) = p_\text{both}.
\]

If we assume that $p_\text{loss}$ is small, then the effect of this conditioning is negligible. For a loss scenario $l$, let $g_2(l)$ denote the number of two-qubit gates where both participating qubits were lost. Then the probability of scenario $l$ with readout $r = \mathcal{R}(l)$ is approximately (discounting the above conditioning on previous losses not occurring) 
\begin{equation}\label{eq:pl}
P(l) \approx p_\text{both}^{g_2(l)} p_\text{single}^{|r|-2g_2(l)} 
= (p_\text{loss} \eta)^{g_2(l)} \left( p_\text{loss} \frac{1 - \eta}{2} \right)^{|r|-2g_2(l)}.
\end{equation}

\subsection{Applying the Pauli Envelope}

When a qubit $q$ is lost at instruction $t$, we model this using the Pauli envelope of the loss scenario $\{(q,t)\}$. Specifically, we insert fully depolarizing noise channels (Stim instruction \texttt{DEPOLARIZE1(0.75)} ) at the relevant locations.\\

The \texttt{DEPOLARIZE1(0.75)} channel samples uniformly from $\{I, X, Y, Z\}$ with probability 0.25 each. For a lost qubit $q$, these channels are inserted at three types of locations:
\begin{enumerate}
\item Immediately before instruction $t$ (the loss location)
\item At role-change boundaries where $q$ switches between control and target in CX gates
\item Before measurement of $q$
\end{enumerate}

The role-change insertions correspond to Hadamard-gate boundaries in the decomposition of CX gates as $H(\text{target}) - CZ - H(\text{target})$. When a lost qubit switches roles, we must re-randomize its Pauli state (by inserting depolarizing noise) to account for the implicit Hadamard.\\

For a loss scenario $l$, let $D(l)$ denote its Pauli envelope DEM. This is obtained by removing all Pauli noise mechanisms from the circuit $\mathcal{C}$, applying the envelope construction to each lost qubit, and compiling the resulting circuit into a detector error model. This DEM specifies a collection of detector-observable patterns $\{(\mathcal{D}_{l,i}, \mathcal{O}_{l,i})\}_{i=1}^{j_l} \subseteq \mathbb{F}_2^{n_d} \times \mathbb{F}_2^{n_\ell}$ and associated probabilities $\{p_{l,i}\}_{i=1}^{j_l}$.\\

An important property of the Pauli envelope is that for most single-qubit loss events, the envelope span equals the set of loss patterns produced by gate stripping. Let the envelope rank of the loss scenario $l$ be defined as 
\[
\text{rank}_e(l) = \dim \text{span } S(E(l), \emptyset)
\]

We may similarly define the loss pattern rank 
\[
\text{rank}_\ell(l) = \dim \text{span } S(\emptyset, l). 
\]

Theorem \ref{thm:pauli_envelope_superset} guarantees that $\text{rank}_e(l) \geq \text{rank}_\ell(l)$. \\

In extensive testing on the Gross [[144,12,12]] code with two syndrome extraction rounds (3312 distinct single-qubit loss events), we found that $S(E(l), \emptyset)$ and $S(\emptyset, l)$ were subspaces for each single-qubit loss event $l$, and that the distribution of detector-observable patterns in each case was uniform. Furthermore, in 66\% of cases $\text{rank}_e(l) = \text{rank}_\ell(l)$, and in the remaining 34\% of cases the envelope rank exceeds the loss pattern rank by exactly one. There were no cases with two or more extra dimensions. Thus, the Pauli envelope contains at most one extra detector-observable pattern generator, and in all cases tested, this extra generator contained no observable flips. See Appendix~\ref{sec:envelope_vs_stripping} for validation that the Pauli envelope strategy for simulating and decoding loss yields logical error rates consistent with the gate-stripping strategy (corresponding to using the patterns in $S(\emptyset, l)$). The Pauli envelope strategy was used because enumerating the loss patterns in $S(\emptyset, l)$, corresponding to gate-stripping, requires sampling from the circuit many times to generate a DEM, whereas a DEM is easily automatically generated for Pauli envelope noise.\\

\section{Decoding Atom Loss using Linear Combinations of DEMs}

After syndrome extraction is completed, a decoder must use the detector syndrome $s \in \mathbb{F}_2^{n_d}$ and loss readout $r \subseteq Q$ to predict the logical observable flip. The challenge is that a loss readout does not give timing information about where losses may have occurred; a single loss readout may correspond to many different loss scenarios. To overcome this difficulty, Baranes et al. \cite{baranes2026} propose constructing a new decoding hypergraph (or DEM) specific to each loss readout. This DEM is constructed by taking linear combinations of DEMs specific to certain loss scenarios. First we will discuss how to construct linear combinations of DEMs, then we will discuss the approach by Baranes et al. (here termed ``Uniform Lifecycle Mixture Decoding''), then we will discuss a new approach based on weighting by conditional probability of loss scenarios and pruning low-probability scenarios.\\

\subsection{Linear Combinations of DEMs}

Consider a Clifford circuit $\mathcal{C}$ with $n_d$ detectors and $n_\ell$ observables. Recall that a \emph{detector error model} (DEM) consists of the linear map between error mechanisms and detector/observable patterns in $\mathbb{F}_2^{n_d} \times \mathbb{F}_2^{n_\ell}$, along with the probabilities associated with each fault. Pauli faults yield error mechanisms and loss events yield multiple error mechanisms (corresponding to multiple possible detector/observable patterns). When multiple error mechanisms with probabilities $\{p_1, \ldots, p_n\}$ yield the same detector-observable pattern, they are indistinguishable for decoding purposes. Thus they may be combined into a single error mechanism. Since detector and observable flips combine by binary sum (XOR), the probability of the summed error mechanism is given by the probability that an odd number of the summand error mechanisms fires, assuming they are independent:
\begin{equation}\label{eq:sum_dems}
p_\text{sum} = \sum_{\substack{S \subseteq [n] \\ |S| \text{ odd}}}
 \prod_{i \in S} p_i \prod_{i \notin S} (1-p_i)
\end{equation}
where $[n] = \{1, \ldots, n\}$.\\

We construct a sum of DEMs by taking the union of all detector-observable patterns. If the same detector-observable pattern is present in multiple of the summand DEMs, we compute the probability in the sum DEM using Eq.~\eqref{eq:sum_dems}. For small probabilities, we may use a truncation of Eq.~\eqref{eq:sum_dems} which considers the probability that exactly one or three of the mechanisms fires using 
\[ 
p_\text{one} = \sum_{i =1}^n p_i \prod_{j\neq i} (1-p_j) = \left[\prod_j(1-p_j)\right] \sum_i \frac{p_i}{1-p_i}, \qquad p_\text{three} = \sum_{i,j,k \in [n]} p_i p_j p_k \prod_{m \neq i,j,k} (1-p_m)
\]
Then $p_\text{sum}$ can be approximated using 
\[
p_\text{sum} = p_\text{one} + p_\text{three} + O(p^5), \qquad p_\text{sum} = p_\text{one} + O(p^3)
\]
where $p := \max_{i\in[n]} p_i$.\\

Multiplying a DEM by a scalar $w \in [0,1]$ scales each mechanism probability: if $D$ has mechanisms $\{(\mathcal{D}_i, \mathcal{O}_i, p_i)\}$, then $wD$ has mechanisms $\{(\mathcal{D}_i, \mathcal{O}_i, wp_i)\}$.\\

\subsection{Uniform Lifecycle Mixture Decoding} \label{sec:UM}

For a loss scenario $l$, recall that $D(l)$ denotes its Pauli envelope DEM, obtained by removing all Pauli noise mechanisms from the circuit $\mathcal{C}$, applying the envelope construction to each lost qubit, and compiling the resulting circuit into a detector error model. In the uniform lifecycle mixture decoding approach, we first precompute a \emph{lifecycle-mixture DEM} for each qubit $q$ by averaging over all possible locations where qubit $q$ could be lost \cite{baranes2026}. In the approach we are using, the probability of $q$ being lost at any given gate (conditional upon not already being lost) is 
\[
p_\text{single} + p_\text{both} = p_\text{loss}(1-\eta)/2 + p_\text{loss}\eta = p_\text{loss}(1 + \eta)/2,
\]
so the lifecycle-mixture DEM for a qubit $q$ which has $|T_q|$ possible loss times is just 
\begin{equation}\label{eq:lifecycle_mixture_dem}
D_\text{lifecycle}(q) = \sum_{t \in T_q} p(\{ (q, t)\} | q \in r) D(\{(q, t)\}) \approx \frac{1}{|T_q|} \sum_{t \in T_q} D(\{(q, t)\}).
\end{equation}

Given a loss readout $r$, we then combine the base Pauli DEM $D_0$ with the sum of all of the relevant lifecycle DEMs \cite{baranes2026}.
\[ 
D_\text{ULM}(r) = D_0 + \sum_{q \in r} D_\text{lifecycle}(q). 
\]

This DEM is then ingested by a decoder. \\

Note that Baranes et al. construct $D(l)$ by computing all of the possible detector/observable patterns resulting from a loss. We compute $D(l)$ via the Pauli envelope of the loss, which is easier to implement. The two approaches are identical in most cases, and where they differ it is because the latter includes an extra detector-only mechanism.\\

This decoding approach does not use DEMs based on multi-loss scenarios, and does not take advantage of any correlations between qubit loss times. However, it still performs fairly well.\\

\subsection{Weighting by Conditional Probability}

The key challenge in decoding atom loss is that $r$ does not specify when each qubit was lost. Multiple scenarios in $\mathcal{R}^{-1}(r)$ are consistent with the readout. The uniform lifecycle mixture approach deals with this by averaging over qubit lifetimes, but one may also consider a weighted approach which averages over all scenarios consistent with the readout, weighting by conditional probability. For a given readout $r$, we can compute 
\[
P(l \mid r) = \frac{P(l)}{\sum_{l' \in \mathcal{R}^{-1}(r)} P(l')}
\]

and then take 
\[ 
D_\text{WCP}(r) = \sum_{l \in \mathcal{R}^{-1}(r)} P(l \mid r) D(l)
\]
as the loss contribution to the final DEM.\\

For small readouts, we can enumerate all $\prod_{q \in r} |T_q|$ scenarios and compute their probabilities. However, this becomes intractable as $|r|$ grows. A pruning strategy discards scenarios with $P(l \mid r) < P_{\max}/F$ where $P_{\max} = \max_{l'} P(l' \mid r)$ and $F$ is a pruning factor (typically 100). The remaining scenarios are combined:
\[
D_\text{Pruned-WCP}(r) = \sum_{l \in \text{kept}} P(l \mid r) \cdot D(l).
\]\\

While this provides more accurate weighting than uniform mixing, it still requires enumerating exponentially many scenarios before pruning can be applied. This quickly becomes computationally infeasible for large readouts and long circuits.\\

On the Gross code with two syndrome rounds, $|T_q| = 12$ for a data qubit and $6$ for an ancilla qubit, so with a readout of size $|r| = 8$ consisting of data qubits, the number of scenarios would be  $12^8 \approx 4 \times 10^8$. Most of these scenarios have very low conditional probabilities $P(l \mid r)$, but enumerating each and computing $P(l | r)$ is still required.\\

\subsection{Maximal-Matching Mixture Decoding} \label{sec:MMM}

A useful and computationally tractable approximation to $D_\text{WCP}(r)$ can be obtained by taking advantage of correlated loss in the noise model. Recall that $g_2(l)$ is the number of two-qubit gates where both participating qubits were lost, and the probability of scenario $l$ with readout $r = \mathcal{R}(l)$ is approximately (discounting the above conditioning on previous losses not occurring) given by Eq. \eqref{eq:pl}, which simplifies to
\[
P(l) \approx \left( p_\text{loss} \frac{1-\eta}{2} \right)^{|r|} \left( \frac{4 \eta}{p_\text{loss}(1-\eta)^2} \right)^{g_2(l)} 
\]

The first factor is constant for a given readout, but the second factor grows exponentially with $g_2(l)$. At typical parameters ($p \approx 0.001$, $\eta \approx 0.5$), this factor is approximately $8 \times 10^{3}$ per correlated pair. Scenarios with maximal $g_2(l)$ therefore dominate the probability mass. We can exploit this by enumerating only the high-probability scenarios rather than the full preimage.\\

Given a readout $r$, construct a graph $G_r$ with vertex set $r$ and an edge $(a,b,t)$ whenever qubits $a, b \in r$ share a two-qubit gate at instruction $t$. ($G_r$ could be a multigraph if two qubits share more than one gate; but in practice this does not occur since qubits are reindexed after a reset). A matching $m \subseteq E(G_r)$ is a set of vertex-disjoint edges. Let $V(m)$ denote the endpoints of edges in $m$, and let $U(m) := r \setminus V(m)$ denote the unmatched vertices. Let $M_{\max}$ be the set of all maximum-cardinality matchings of $G_r$.\\

To enumerate $M_{\max}$, we first split $G_r$ into its connected components. A maximum matching of $G_r$ is exactly a union of one maximum matching from each component, so it suffices to enumerate each component's maximum matchings independently and take the Cartesian product. Within a component we use a branch-and-bound depth-first search: pick an unprocessed edge and branch on either including it (removing both endpoints and all incident edges) or excluding it, pruning any branch whose remaining edges cannot reach the best matching size found so far. Because two lost qubits can only be joined by an edge of $G_r$ if they share a gate, each component is confined to a single teleportation window.  With a teleportation period of $\mathrm{tp} \leq 2$, no component of $G_r$ has more than $\sim\!12$ edges regardless of $|r|$ or $n_r$, so the DFS is exhaustive and cheap. When the Cartesian product exceeds a fixed cap (which only occurs in experiments with $\mathrm{tp} = n_r \gg 2$) we select an evenly-spaced representative subset using a Hopcroft--Karp maximum-matching pass starting from eight independently randomized edge orderings. \\

Once we have found all $m \in M_{\max}$, we have to account for the probability of all loss scenarios corresponding to it. For each maximal matching $m$, we define the cell $C(m)$ as the set of scenarios that place each matched pair $(a,b,t) \in m$ at instruction $t$ and allow each unmatched qubit $q \in U(m)$ to choose any $t_q \in T_q$. Each loss scenario in $\cup_{m \in M_{\max}} C(m)$ has approximately the same probability according to Eq.~\eqref{eq:pl} since it has the same number of two-qubit gates and the same readout. Therefore, the probability of a whole cell $C(m)$ is proportional to its cardinality, which is the number of ways to choose loss times for the unmatched qubits in $U(m)$, i.e.
\[
W_m = |C(m)| = \prod_{q \in U(m)} |T_q|.
\]
For example, in the Gross code with 2 rounds, a matching that leaves a data qubit ($|T_q| = 12$) unmatched instead of an ancilla qubit ($|T_q| = 6$) gets twice the weight.\\

Scenarios not in some $C(m)$ for $m \in M_{\max}$ have negligible probability and can be discarded. With all scenarios not in any $C(m)$ discarded, we can approximate the conditional probability of each cell as $W_m / \sum_{m'} W_{m'}$ and collapse the weighted-conditional-probability mixture from $\prod_{q \in r} |T_q|$ terms to $|M_{\max}|$ terms:
\begin{equation}\label{eq:wcp-maximal}
D_\text{WCP}(r) \approx D_\text{WCP-maximal}(r) = \sum_{m \in M_{\max}} \frac{W_m}{\sum_{m'} W_{m'}} \langle D(l) \rangle_{l \in C(m)}.
\end{equation}
where  $\langle D(l) \rangle_{l \in C(m)}$ is the averaged DEM over cell $C(m)$. \\

The cell average DEM can be decomposed over individual qubits. For $l \in C(m)$, the Pauli envelope DEM decomposes as
\[
D(l) = \sum_{(a,b,t) \in m} D(\{(a,t), (b,t)\}) + \sum_{q \in U(m)} D(\{(q, t_q)\})
\]
where the first sum corresponds to correlated pairs and the second to unmatched qubits. Averaging over all choices of $\{t_q : q \in U(m)\}$ gives
\[
\langle D(l) \rangle_{l \in C(m)} = \sum_{(a,b,t) \in m} D(\{(a,t), (b,t)\}) + \sum_{q \in U(m)} D_\text{lifecycle}(q),
\]
using the lifecycle-mixture DEM for qubit $q$, defined in Eq.~\eqref{eq:lifecycle_mixture_dem}. Thus Eq.~\eqref{eq:wcp-maximal} simplifies to
\[
D_\text{WCP-maximal}(r) =  \sum_{m \in M_{\max}} \frac{W_m}{\sum_{m'} W_{m'}} \left[\sum_{(a,b,t) \in m} D(\{(a,t), (b,t)\}) + \sum_{q \in U(m)} D_\text{lifecycle}(q)\right] \approx D_\text{WCP}(r)
\]

To pass to a decoder, we combine with the contribution from Pauli errors to yield 
\[ 
D_\text{MMM}(r) := D_0 + D_\text{WCP-maximal}(r).
\]

This reduces to the $D_\text{ULM}$ decoder when $M_{\max} = \{\emptyset\}$ (no shared gates). When exactly one pair shares a gate, it equals $D_\text{ULM}$ on the other qubits plus the correct joint envelope on the pair. Thus, the maximal-matching mixture can be viewed as a correction to $D_\text{ULM}$ whenever the readout indicates two qubits were most likely lost together. Furthermore, enumerating $M_{\max}$ is inexpensive for typical readouts. \\ 

\section{Methods}\label{sec:methods}

\subsection{Test Circuit Generation Using Stim}

We generate syndrome extraction circuits for the Gross code using the Stim stabilizer circuit simulator \cite{gidney2021stim}. The circuit structure includes initialization of data and ancilla qubits, multiple rounds of noisy syndrome extraction with two-qubit gates, measurements with detector annotations, and background noise channels. After the noisy syndrome extraction rounds, a noiseless syndrome extraction round is included to make sure that the corrections proposed by the decoder can return the circuit state to the codespace. \\

For codes with multiple syndrome extraction rounds, ancilla qubits are reset between rounds. To simplify the loss tracking, we first apply an unrolling transformation that assigns each qubit lifetime a unique qubit index and removes SWAPs in the circuit. After this transformation, each qubit index corresponds to exactly one physical qubit lifetime, eliminating the need to track reset boundaries during loss analysis. Using this method, each loss readout only needs to refer to qubit indices, and there will be no ambiguity about which physical qubit lifetime is being referred to. \\

We confine loss tracking to the portion of the circuit where noisy syndrome extraction takes place. In particular, the encoder, noiseless syndrome extraction round, and measurement of logical observables are assumed to not be subject to loss or Pauli noise. \\

\subsection{Hierarchical Sampling Procedure}

The hierarchical sampling procedure generates syndrome and loss data for each sample. Given a certain number $N$ of samples to take, we first sample $N$ loss scenarios. These scenarios are grouped if there are any duplicates (which was unlikely in our circuits for the noise levels we tested, since there are hundreds of qubits which may be lost). Suppose that there are $M$ unique loss scenarios, each corresponding to $n_i \geq 1$ samples (where $1 \leq i \leq M$). Then for each unique loss scenario, we obtain the DEM corresponding to a circuit containing the original Pauli errors mechanisms along with the Pauli envelope of the loss added to the circuit. We use this DEM to sample $n_i$ error vectors (via Stim's compiled DEM sampler \cite{gidney2021stim}). 

\subsubsection{Sampling Loss Scenarios}

The sampling procedure first requires enumerating all possible loss mechanisms. For each qubit $q$, we identify its life cycle events: the set of all two-qubit gates involving $q$. We enumerate all loss mechanisms across all qubits:
\begin{itemize}
\item Single-qubit loss at two-qubit gate $(a,b,t)$: two mechanisms with probability $p_{\text{single}}$ each
\item Correlated two-qubit loss at gate $(a,b,t)$: one mechanism with probability $p_{\text{both}}$
\end{itemize}

To sample a loss scenario, we sample loss mechanisms as independent Bernoulli trials. If multiple mechanisms affect the same qubit, we retain only the earliest loss time for that qubit. This yields a loss scenario $l = \{(q_1, t_1), (q_2, t_2), \ldots\}$ and its corresponding readout $r = \mathcal{R}(l)$.\\

\subsubsection{Software Optimization}

Circuits and DEMs for loss scenarios are constructed on-demand using lazy evaluation, both in the sampling and decoding process. DEMs are compiled only when a particular scenario is first encountered, then cached for reuse. This dramatically reduces memory requirements compared to pre-generating all possible scenarios.\\

Because the time cost of sampling and of decoding is dominated by loss DEM generation, we implement an option to generate loss scenarios in batches. Given a batch size $B$, the sampler would then generate $N/B$ loss scenarios. Then, for the resulting $M$ unique loss scenarios, each corresponding to $n_i \geq 1$ samples (where $1 \leq i \leq M$ and $\sum_{i=1}^M n_i = N/B$), we sample $B n_i$ error vectors from the corresponding Pauli envelope loss DEM. This strategy reduces the overhead of building DEMs by roughly a factor of $B$ but does not change the per-sample cost of sampling or decoding. Batching with $B > 1$ offers a significant speed up with a decrease in memory usage, and leaves the per-sample marginal distribution unchanged. However, it makes the samples within each batch correlated, thereby increasing the width of confidence intervals. The experiments in this paper use $B = 1$. \\

\section{Results}\label{sec:results}

In our experiments, we compare logical error rates for three decoding strategies, which all create a DEM for each loss readout and decode using Relay-BP with parameters \texttt{gamma0=0.65}, \texttt{pre\_iter=80}, \texttt{num\_sets=100}, \texttt{set\_max\_iter=60}, \texttt{gamma\_dist\_interval=(-0.24, 0.66)}, \texttt{stop\_nconv=5}:
\begin{itemize}
  \item \textbf{Loss-unaware (LU)} ignores the loss readout entirely
    and decodes using the base Pauli DEM. This is used as a baseline for comparison; and it is a somewhat pessimistic baseline because it assumes loss is completely unheralded.
  \item \textbf{Uniform-mixture (UM)} is loss-aware: for each lost
    qubit it averages the decoding model over the possible times that
    qubit could have been lost, treating each lost qubit
    independently. This is the approach taken by Baranes et al. \cite{baranes2026} and outlined in Section \ref{sec:UM}.
  \item \textbf{Maximal-matching-mixture (MMM)} is loss-aware and weights loss scenarios by conditional probability; discarding all scenarios which are not maximally-matched (in the sense of including as many two-qubit losses as possible) because these scenarios are much more unlikely. This approach is outlined in Section \ref{sec:MMM}. 
\end{itemize}
We use $n_r$ to represent the number of noisy syndrome extraction rounds, $p$ to represent the total two-qubit gate failure probability (of which a proportion $\alpha$ is loss and a proportion $1 - \alpha$ is depolarizing noise), and $\eta$ to represent the probability that a gate loses both qubits when it experiences a loss event (the pairing probability). After every $\mathrm{tp}$ syndrome extraction rounds (where $n_r$ is divisible by $\mathrm{tp}$), each data qubit is teleported onto a fresh ancilla by the gadget shown in Fig.~\ref{fig:teleportation_gadget}. \\ 

Throughout, the \emph{loss-aware advantage} means the ratio of the loss-unaware decoder's logical error rate to a loss-aware decoder's, i.e.\ how many times better the loss-aware decoder is. Logical error rates are reported per syndrome-extraction round (see Appendix~\ref{sec:LER}) and error bars use binomial Wilson $95\%$ confidence intervals. \\

\subsection{Loss-aware Advantage vs Loss Fraction}

\begin{figure}[H]
  \centering
  \includegraphics[width=\linewidth]
    {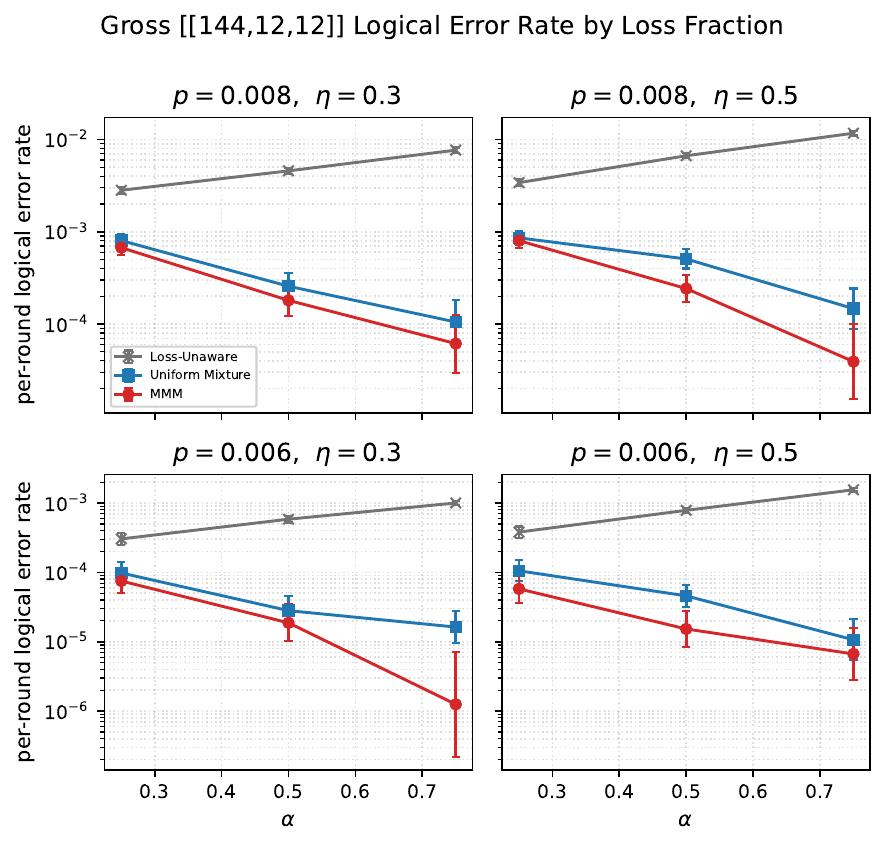}
  \caption{Per-round logical error rate versus loss fraction $\alpha$
  (columns: pairing probability $\eta$; rows: physical error rate
  $p$). Grey: loss-unaware. Blue: uniform-mixture. Red:
  maximal-matching-mixture. As more of the noise budget becomes
  heralded loss, the loss-aware decoders improve relative to the
  loss-unaware baseline; MMM's additional gain over UM appears strongest where
  paired losses are common (large $\alpha$, large $\eta$).}
  \label{fig:ler-vs-lf}
\end{figure}

Figure~\ref{fig:ler-vs-lf} compares logical error rates of our three decoding strategies at varying loss fractions and pairing probabilities, with physical error rates $p{=}0.008, 0.006$, $n_r{=}2$ noisy SE rounds, and $\mathrm{tp}{=}2$ (one teleportation at the end of the circuit). The loss fraction is swept over $\alpha\in\{0.25,0.5,0.75\}$ and the pairing probability is swept over $\eta\in\{0.3,0.5\}$.   The loss-aware advantage grows steeply with~$\alpha$. As $\alpha$ increases, loss-unaware performance degrades (since Pauli errors are traded for loss and heralding information is unused) and loss-aware performance improves (since unheralded Pauli errors are traded for heralded loss). At $p = 0.008$, MMM is about $4\times$ better than LU at $\alpha{=}0.25$, about $25\times$ at $\alpha{=}0.5$, and $125$--$300\times$ at $\alpha{=}0.75$. MMM's additional edge over UM is largest where paired losses are common: at $\alpha{=}0.75$ it grows from $1.7\times$ ($\eta{=}0.3, p=0.008$) to $3.7\times$ ($\eta{=}0.5, p=0.008$); at $\alpha{=}0.25$ the two loss-aware decoders are nearly equal, since with mostly depolarizing noise there is less pairing structure to exploit.\\

\begin{figure}[H]
  \centering
  \includegraphics[width=\linewidth]
    {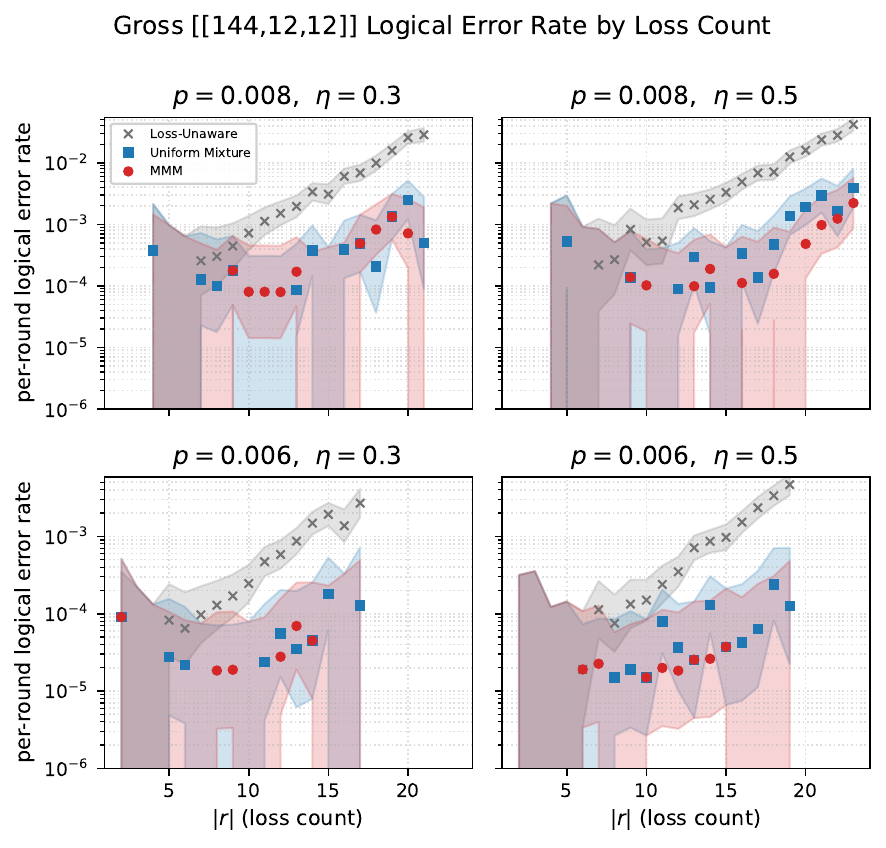}
  \caption{Per-round logical error rate versus the number of lost qubits $|r|$ in a sample (rows: $p{=}0.008, 0.006$, columns: $\eta=0.3, 0.5$, $\alpha{=}0.5$.). Wilson 95\% confidence intervals are shown in shading. The loss-unaware decoder's logical error rate grows roughly geometrically in $|r|$; the loss-aware decoders' logical error rate grows much more slowly, so the loss-aware advantage is concentrated in the high-$|r|$ tail of the loss-count distribution. Values of $|r|$ which represent less than $1/100$ of the probability mass are excluded, see Figure~\ref{fig:R_histogram} for the distribution of loss counts across these four scenarios.}
  \label{fig:ler-by-R}
\end{figure}

Figure~\ref{fig:ler-by-R} analyzes the same data by the number of lost qubits (or size of loss readout, $|r|$). After a decrease for small values of $|r|$ which are unlikely, the loss-unaware decoder's error rate rises steeply for each additional lost qubit, while the logical error rates for both loss-aware decoders rise much more slowly. Essentially all of LU's failures come from the above-average-loss shots that the loss-aware decoders handle correctly. \\

\begin{figure}[H]
  \centering
  \includegraphics[width=\linewidth]
    {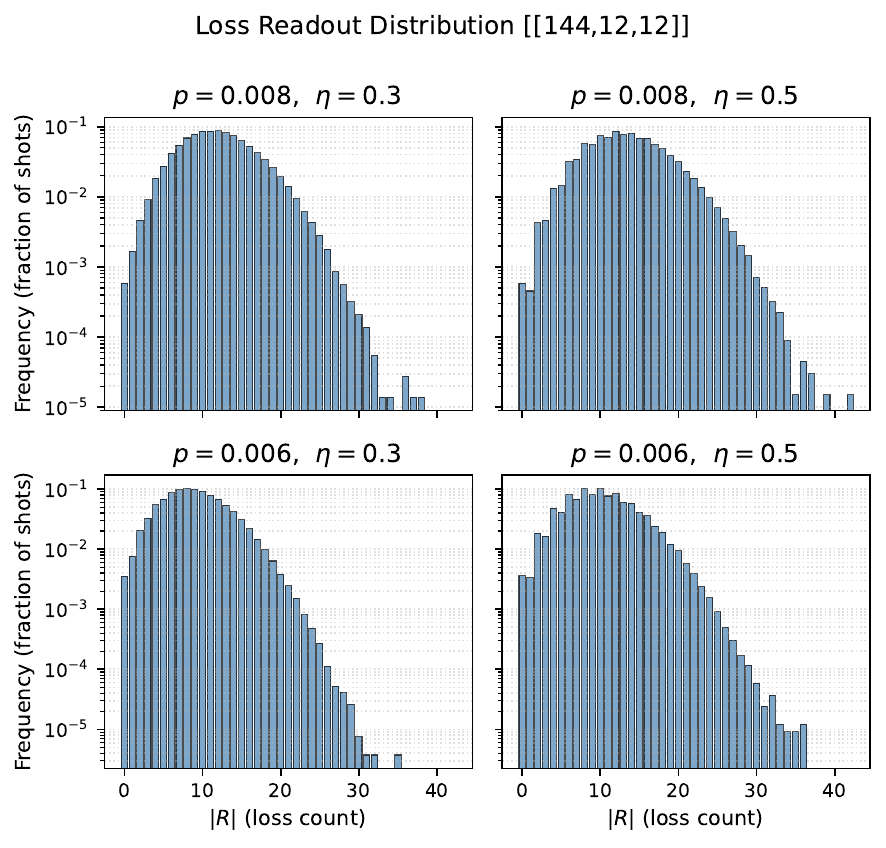}
  \caption{Distribution of loss counts in the experiment from Figure~\ref{fig:ler-by-R} (rows: $p{=}0.008, 0.006$, columns: $\eta=0.3, 0.5$, $\alpha{=}0.5$.). From left to right then top to bottom, the means are 12.04, 13.49, 9.07, 10.17. }
  \label{fig:R_histogram}
\end{figure}

\subsection{Error rates and Loss-aware Advantage by Teleportation Period}

\begin{figure}[H]
  \centering
  \includegraphics[width=\linewidth]
    {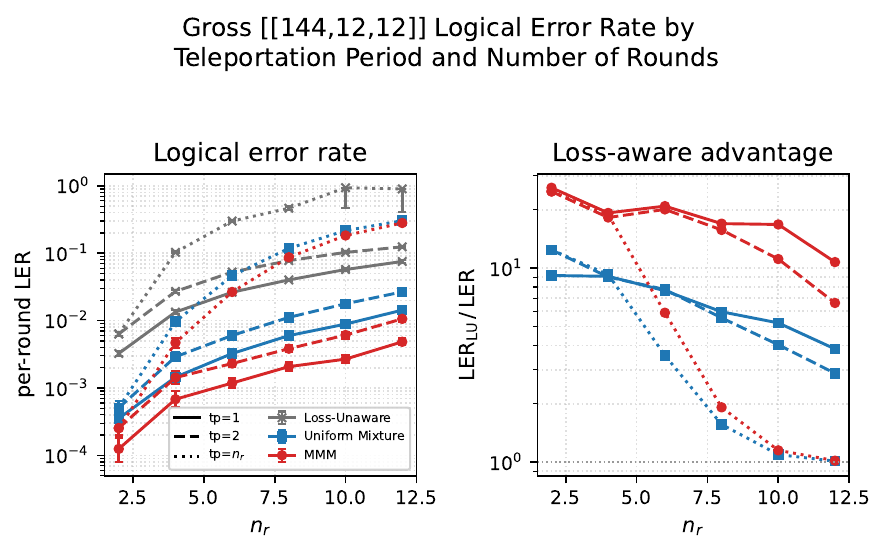}
  \caption{Per-round logical error rate versus number of rounds $n_r$, at three teleportation periods (panels). With teleportation every one or two rounds, the gap between loss-aware and loss-unaware decoders is relatively flat in $n_r$; with only a single teleportation at the end, the gap closes as $n_r$ grows and MMM approaches UM.}
  \label{fig:teleport-vs-rounds}
\end{figure}

Figure~\ref{fig:teleport-vs-rounds} varies the number of rounds from $n_r = 2$ to $n_r = 12$ in increments of two, with the teleportation period chosen from  $\mathrm{tp}\in\{1,2,n_r\}$ at $p{=}0.008$, $\alpha{=}\eta{=}0.5$.
When data qubits are teleported at least every other round ($\mathrm{tp}\le 2$), the loss-aware advantage is essentially independent of $n_r$: MMM stays $17$--$25\times$ better than LU and $2$--$3\times$ better than UM out to $n_r{=}12$. When there is only a single teleportation at the end ($\mathrm{tp}{=}n_r$), the loss-aware advantage collapses as $n_r$ grows: MMM's edge over LU falls from $25\times$ at $n_r{=}2$ to $3$--$5\times$ at $n_r{\ge}8$, as the logical error rate of LU saturates to $\approx 1$ for very high $n_r$. Therefore, periodic teleportation or other loss detection capability is key to implementing loss-tolerant quantum circuits. \\ 

\subsection{Teleportation Noise Ablation}

\begin{figure}[H]
  \centering
  \includegraphics[width=\linewidth]
    {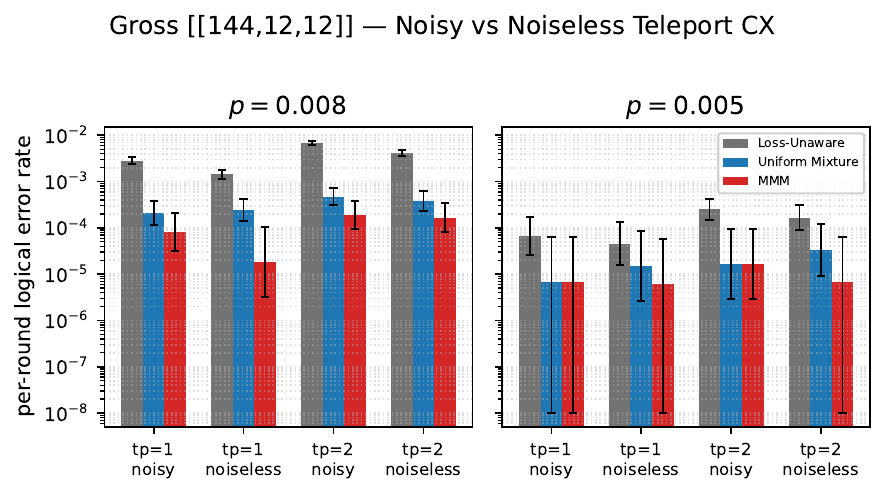}
  \caption{Per-round logical error rate at $n_r{=}2$ for teleportation period $\mathrm{tp}\in\{1,2\}$ with the teleportation CNOT either physical (subject to loss and depolarization) or idealized (perfect). Left panel $p{=}0.008$, right $p{=}0.005$. With a physical CNOT, teleporting every round is still a net ${\sim}2\times$ improvement over teleporting once.}
  \label{fig:tp-noise-ablation}
\end{figure}

Teleporting more often shortens qubit lifetimes but adds a noisy CNOT each time. Figure~\ref{fig:tp-noise-ablation} attempts to separate these two effects at $n_r{=}2$, $p{=}0.008$, by comparing runs where the teleportation CNOT is physically noisy (subject to loss and Pauli depolarizing noise) to runs where it is noiseless (idealized). For MMM, the physical CNOT costs $1.2\times$ at $\mathrm{tp}{=}2$ (one teleportation) and about $4.6\times$ at $\mathrm{tp}{=}1$ (two). Halving the loss window (going from $\mathrm{tp}{=}2$ to $\mathrm{tp}{=}1$) is worth about $8\times$ with an idealized CNOT but only about $2\times$ with a physical one-- the noise from the extra gate consumes most of the gain. Nevertheless,  teleporting after every round, even with a fully physical teleportation gate, still outperforms by a factor of $2\times$. 

\subsection{Logical Error Rate Scaling}

\begin{figure}[H]
  \centering
  \includegraphics[width=0.7\linewidth]
    {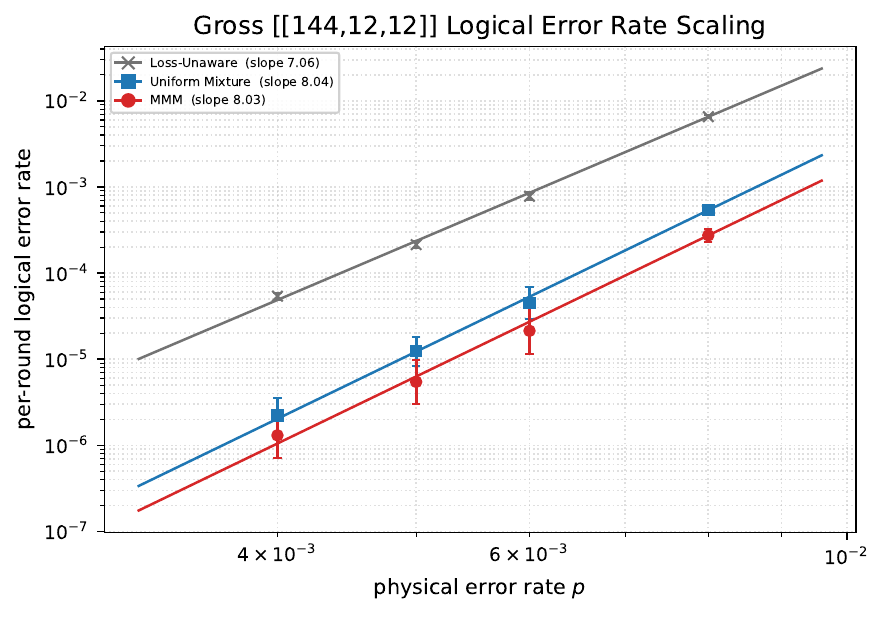}
  \caption{Per-round logical error rate versus physical error rate $p$ (log-log; $\alpha{=}\eta{=}0.5$, $n_r{=}2$). Wilson 95\% confidence intervals are shown. Solid lines are power-law fits. The loss-aware decoders have a steeper slope (8 vs 7) than the loss-unaware baseline, so their advantage widens as physical error rate $p$ decreases.}
  \label{fig:ler-scaling}
\end{figure}

Figure~\ref{fig:ler-scaling} plots the per-round logical error rate against $p\in\{4,5,6,8\}{\times}10^{-3}$ on log-log axes ($\alpha{=}\eta{=}0.5$, $n_r{=}2$, $\mathrm{tp}{=}2$). The loss-unaware slope is $7$, and the slopes for the loss-aware decoders are $8$. Thus, the loss-aware advantage grows as $p$ decreases: MMM's improvement over LU rises from about $24\times$ at $p{=}0.008$ to about $41\times$ at $p{=}0.004$. MMM's edge over UM stays roughly flat at ${\sim}2\times$ across the range.

\section{Conclusion and Further Work} 

Building circuits and decoders which are tolerant to delayed-heralded atom loss is key to building scalable neutral atom quantum processors. This work presents a {maximal-matching-mixture} (MMM) decoder, which uses a loss readout to build a decoding hypergraph as a weighted mixture over exactly the loss scenarios consistent with the readout.  Losses at a two-qubit gate are strongly correlated, and MMM takes advantage of this correlation to intelligently prune unlikely loss scenarios. On a $[[144,12,12]]$ Gross code syndrome extraction circuit, MMM lowers the logical error rate by ${\sim}25\times$ relative to a loss-unaware decoder and ${\sim}2\times$ relative to a loss-aware uniform-mixture baseline \cite{baranes2026}. 

There are several interesting directions for further work. Liu et al.~\cite{liu2026} showed that adding a loss exclusivity constraint to a decoding problem represented as an integer linear program (ILP) significantly improved logical error rate scaling; at the cost of having to use a general ILP solver. Efficiently implementing such a loss exclusivity constraint in a more standard decoding problem (e.g. with a decoding hypergraph) is an open problem. \\ 

The maximal-matching mixture treats all maximally-matched scenarios as a priori comparable, weighting them only by their unmatched-qubit degrees of freedom. Belief propagation based decoders already operate with a probabilistic distribution of errors; perhaps there is some way of passing the matching probability distribution directly to belief propagation, rather than collapsing it to a single mixed DEM. Alternatively, one could take an iterative approach; starting with a mixture DEM but then selecting the loss scenario with the highest posterior probability given the output of belief propagation, then decoding using that specific loss scenario. \\

The error model for loss in this work does not account for the possibility of combined loss and Pauli errors, even though it has been shown that if one qubit is lost during a Rydberg gate, its two-qubit partner is subject to strong biased Pauli noise \cite{correlatedloss}. Testing this biased, correlated, combination Pauli and loss noise is a key follow-up to this work. \\

Finally, nothing in the construction is specific to the Gross code or to bare-ancilla syndrome extraction--- extending the experiments to other qLDPC codes, to other types of syndrome extraction, and to logical operations rather than memory, would provide important tests of the loss-aware advantage. \\

\section*{AI Disclosure}

In preparing this work, Claude was used to assist with implementation, testing, debugging, and optimization of software. It was also used to aggregate results, assist in creating visualizations, and assist with debugging and proofreading the draft in \LaTeX.

\bibliographystyle{plain}
\bibliography{loss}

\appendix

\section{Logical Errors} \label{sec:LER}

Given a quantum error-correcting code which encodes $k$ logical qubits into $n$ physical qubits, a logical observable is an operator which acts on the $2^k$-dimensional logical subspace (analogous to how physical observables act on physical qubits). If the code is a stabilizer code, then logical observables are those that commute with all the stabilizers of the code, while not themselves being stabilizers. Measuring a logical observable extracts information about the encoded quantum state without collapsing information about other observables.\\

In fault-tolerant quantum computation, a set of qubits is placed into a state within the codespace of a quantum error-correcting code, and the data is contained in the logical observables of the code, which store the data about the $k$ logical qubits. When physical errors occur, a syndrome extraction circuit measures the stabilizers and a classical algorithm called a  decoder tries to find a correction that returns the state to the code space. A \emph{logical error} is a physical error process that implements a nontrivial logical observable even after corrected by a decoder (i.e. the decoder returned to the codespace but not to the correct state, and a logical observable has flipped).\\

For a syndrome extraction circuit, the logical error rate, or the probability of a logical error under a certain noise model, is a key QEC metric. If $n_r > 1$ rounds of syndrome extraction are performed, then the per-round logical error rate is calculated as 
\[
LER_{round} = 1 - (1 - LER)^{1/n_r}.
\]\\

\section{Faithfulness of the Pauli-Envelope Surrogate for Atom Loss} \label{sec:envelope_vs_stripping}

This work simulates atom loss by applying the Pauli envelope of the loss as defined in \cite{liu2026}. Rather than stripping gates which do not operate due to loss, the simulator inserts fully depolarizing single-qubit noise channels at the time of loss, where CNOT role changes occur, and where the qubit is measured. The loss circuits passed to the decoder also use the Pauli envelope. To validate that our strategy (the ``Pauli-Envelope surrogate'') obtains similar logical error rates to the (more physically accurate) gate-stripping strategy for loss modeling, we implemented the latter strategy and tested it on our Gross [[144,12,12]] code syndrome extraction circuit with four rounds. \\

Finding the detector/observable patterns resulting from a loss event in a Stim circuit requires some care. We first strip out the gates which a lost qubit participates in after the time it is lost. But generating a detector error model is slightly unintuitive. Because Stim computes detector \emph{flips} compared to a noiseless baseline (called a reference sample), a naive call to generate a DEM on the gate-stripped circuit will only yield a list of detectors which are non-deterministic (the gate-stripped circuit has nothing that Stim considers a noise mechanism). Thus, we must sample detector-observable flips from the gate-stripped circuit \emph{and} the original circuit baseline, each with a common reference sample (usually all zeros). Then, the binary addition (XOR) of the two sets of detector patterns will yield possible detector/observable flip patterns. We must take enough samples to capture all possible detector/observable patterns and to accurately estimate their probability using the sample mean. For this reason, this strategy is more expensive than using the Pauli envelope for generating loss-circuit DEMs. Furthermore, the experiment below shows that using the Pauli envelope yields, at worst, a slightly conservative estimate of the logical error rate of a circuit.\\

For this experiment, we used a physical error rate of $p = 0.008$, loss fraction $\alpha = 0.5$, fraction of losses where both qubits are lost $\eta = 0.5$, and a single teleportation of data qubits at the end of the 4th round. 2000 samples of each configuration were taken using identical loss patterns. Table~\ref{tab:envelope_vs_strip} compares each decoder's logical error rate under this surrogate to its rate under the true gate-stripped circuit.\\

\begin{table}[h]
\centering
\begin{tabular}{llrr}
\hline
Decoder & Envelope LER [95\% CI] & Strip LER [95\% CI] & Ratio \\
\hline
Loss-unaware  & $0.359\;[0.338,0.380]$ & $0.328\;[0.308,0.349]$ & $0.91$ \\
Uniform-mixture & $0.0345\;[0.0274,0.0434]$ & $0.0290\;[0.0225,0.0373]$ & $0.84$ \\
Max-matching-mixture & $0.0145\;[0.0101,0.0207]$ & $0.0120\;[0.0081,0.0178]$ & $0.83$ \\
\hline
\end{tabular}
\caption{Logical error rates, with 95\% binomial Wilson confidence intervals, for the Pauli envelope surrogate and the gate-stripping strategy for modeling loss. Results are shown for the three decoders used in this work.}
\label{tab:envelope_vs_strip}
\end{table}

The results show that logical error rates from the Pauli envelope surrogate are consistently 10-20\% higher than in the gate-stripping strategy, but the Wilson 95\% confidence intervals overlap. This is consistent with the idea that the Pauli envelope sees a superset of the detector/observable patterns possible under an atom loss. Because the decoder sees slightly more noise under the envelope surrogate than under physical loss, LERs under the envelope surrogate are mildly conservative but broadly consistent with those obtained from a gate-stripping strategy. In particular, the loss-aware advantage is preserved: the performance increase of MMM (maximal-matching mixture) over LU (loss-unaware) is about 25x under the surrogate and about 27x under the gate-stripping strategy.

Note that this small validation experiment does not test whether a decoder derived from gate-stripped loss circuits would perform better than an envelope-derived one on the same set of samples. That would be a separate question. Here we compare the performance of decoders on samples generated by the same strategy as the loss circuits passed to the decoder.

\end{document}